\documentclass[aip,preprint,psfig]{revtex4-1}
\usepackage{psfig}

\begin{document}
\include{epsf}

\title{
Inelastic cross sections and rate coefficients for collisions between CO and H$_2$}

\author{Christina Castro}
\affiliation{Department of Physics, Penn State University,
Berks Campus, Reading, PA 19610-6009}

\author{Kyle Doan}
\affiliation{Department of Physics, Penn State University,
Berks Campus, Reading, PA 19610-6009}

\author{Michael Klemka}
\affiliation{Department of Physics, Penn State University,
Berks Campus, Reading, PA 19610-6009}

\author{Robert C. Forrey}
\email{rcf6@psu.edu}
\affiliation{Department of Physics, Penn State University,
Berks Campus, Reading, PA 19610-6009}

\author{B. H. Yang}
\affiliation{Department of Physics and Astronomy and the Center for
Simulational Physics, University of Georgia, Athens, Georgia 30602}

\author{Phillip C. Stancil}
\affiliation{Department of Physics and Astronomy and the Center for
Simulational Physics, University of Georgia, Athens, Georgia 30602}

\author{N. Balakrishnan}
\affiliation{Department of Chemistry, University of Nevada, Las Vegas, NV 89154}

\date{\today}

\begin{abstract}
A five-dimensional coupled states (5D-CS) approximation is used
to compute cross sections and rate coefficients for CO+H$_2$
collisions. The 5D-CS calculations are benchmarked against
accurate six-dimensional close-coupling (6D-CC) calculations 
for transitions between low-lying rovibrational states. 
Good agreement between the two formulations is found for 
collision energies greater than 10 cm$^{-1}$. 
The 5D-CS approximation is then used to compute two separate databases 
which include highly excited states of CO that are beyond the 
practical limitations of the 6D-CC method. The first database
assumes an internally frozen H$_2$ molecule and allows 
rovibrational transitions for $v\le 5$ and $j\le 30$, 
where $v$ and $j$ are the 
%respective 
vibrational and rotational
quantum numbers of the initial state of the CO molecule.
The second database allows H$_2$ rotational transitions
for initial CO states with $v\le 5$ and $j\le 10$. The two databases
are in good agreement with each other for transitions that are
common to both basis sets. Together they provide data for 
astrophysical models which were previously unavailable.
\end{abstract}


\maketitle

\section{Introduction}

H$_2$ and CO are the most abundant molecules in most astrophysical
environments and have been the focus of numerous astrophysical studies 
and observations \cite{ewine}. CO is easily excited by collisions 
with H$_2$ and other species in interstellar gas, and the
resulting emission lines are commonly used to provide 
important diagnostics of gas density and temperature.
When these environments are irradiated with an intense UV field, 
the radiation drives the chemistry and internal level 
populations out of equilibrium. Models which aim to
interpret the emission lines arising from these environments
must account for all mechanisms which can excite or de-excite 
the molecules. Simulation packages have been developed 
\cite{warin,petit,shaw,rollig}
which account for inelastic transitions of both H$_2$ and CO
in various astrophysical environments such as photodissociation
regions (PDRs) which occur between hot H II (ionized hydrogen) 
and cold molecular regions. Observations of star-forming regions
and protoplanetary disks (PPDs) of young stellar objects have shown 
evidence of rovibrational transitions involving states of CO which 
are highly excited. In particular, vibrational transitions have been 
detected in the near infrared by the {\it Infrared Space Observatory}, 
the Gemini Observatory, and the Very Large Telescope (VLT).
A recent VLT survey \cite{brown} of 69 PPDs found that 77\% of the sources 
showed CO vibrational bands, including vibrational level $v$ as high as 4
and rotational level $j$ as high as 32.
CO vibrational bands in the 1-5 micron regime will soon be accessible 
by NASA's {\it James Webb Space Telescope} which will allow 
the warm inner regions of young stellar objects to be probed.
To reliably model these environments requires an extensive set of
state-to-state rate coefficients for rovibrational transitions 
induced by H$_2$ collisions. Many of these rate coefficients
are either unavailable or else are 
%supplied using estimates which are based on methods 
%that are known to be unreliable \cite{scaling}.
estimated with potentially unreliable methods \cite{scaling}.

To provide accurate data for astrophysical models generally requires 
a large and concerted effort between several communities within
atomic and molecular physics. Experimental data are 
sparse but are critical for benchmarking theoretical methods 
\cite{costes,faure,nature}.
Numerical solution of the Schr\"odinger equation offers the
best hope for generating the bulk of the needed data. However,
to make these computations possible, it is often necessary to
invoke decoupling approximations which reduce the
dimensionality of the full scattering problem. Even when
full-dimensional quantum dynamics calculations are feasible,
their accuracy depends on the underlying electronic structure 
calculations and potential energy surface (PES)
which is used as input. Recently, a high-level full-dimensional 
PES was developed for CO+H$_2$ and used in a six-dimensional 
close-coupling (6D-CC) formulation \cite{nature} which is 
essentially an exact treatment of the dynamics.
The calculations were in excellent agreement with experiment
%for the de-excitation of vibrationally excited CO, whereas
for vibrational de-excitation of CO, whereas
prior calculations which consisted of various levels of
approximation varied by more than two orders of magnitude 
\cite{bacic1,bacic2,reid,flower}.
Subsequently, the 6D-CC approach was used to provide additional 
high-quality cross sections for CO+H$_2$ collisions \cite{yang}.
These calculations are computationally demanding and have been
completed for only a small subset of rovibrational levels
needed for the astrophysical models. Nevertheless, these
calculations provide the best benchmarks for transitions
where experimental data are lacking, and may serve as
a foundation for testing approximate schemes which aim
to further extend the available data. 
One such scheme is the coupled states or
centrifugal sudden (CS) approximation which requires significantly
less computational effort due to the decoupling of 
orbital and internal angular momenta. 
A recent study \cite{forrey} compared 6D-CS and 6D-CC
cross sections for CO+H$_2$ and found encouraging 
results for energies above 10 cm$^{-1}$. Additional
decoupling of the ``twist" angle between the two
molecules led to a 5D-CS approximation which
also yielded encouraging results. In the present
work, we compute a large database of cross sections
and rate coefficients using the 5D-CS approximation.
The 5D-CS calculations are compared with existing
6D-CC results in order to assess the expected 
level of accuracy of the approximation.
The database provides a balance between
accuracy and computational efficiency
and is made available in a convenient
format for use in astrophysical models.

\section{Theory}

The quantum mechanical CC and CS formulations for diatom-diatom collisions 
have been given previously \cite{green,alexander,heil}. 
In order to clarify the differences between the 5D and 6D formulations, 
we provide a brief summary of the theory.
The Hamiltonian of the four-atom system may be written
\begin{eqnarray}
H(\vec{r}_1,\vec{r}_2,\vec{R}) = T(\vec{r}_1) + T(\vec{r}_2) + T(\vec{R}) 
+ V(\vec{r}_1,\vec{r}_2,\vec{R}) ,
\label{HAM}
\end{eqnarray}
where $T(\vec{R})$ is a radial kinetic energy term describing 
the center-of-mass motion, $T(\vec{r}_1)$ and $T(\vec{r}_2)$ 
are kinetic energy terms for the diatomic molecules, and 
$V(\vec{r}_1,\vec{r}_2,\vec{R})$ is the potential energy 
of the system. It is convenient to define
\begin{eqnarray}
V(\vec{r}_1,\vec{r}_2,\vec{R}) = U(\vec{r}_1,\vec{r}_2,\vec{R}) + V(\vec{r}_1) + V(\vec{r}_2) ,
\label{Potential}
\end{eqnarray}
where $V(\vec{r}_1)$ and $V(\vec{r}_2)$ are the two-body potential energies 
of the isolated CO and H$_2$ molecules, and $U(\vec{r}_1,\vec{r}_2,\vec{R})$ 
is the four-body interaction potential which vanishes at large separations. 
The 6D Jacobi coordinate system in Figure 1 is used where
$R$ is the distance between the centers-of-mass of the diatomic 
molecules, $\theta_1$ is the angle between $\vec{r}_1$ and $\vec{R}$, 
$\theta_2$ is the angle between $\vec{r}_2$ and $\vec{R}$, and
$\phi$ is the out-of-plane dihedral angle or ``twist" angle.
The interaction potential may be expanded as
\begin{equation}
U(\vec{r}_1,\vec{r}_2,\vec{R})
%=\sum_{\lambda_1,\lambda_2,\lambda_{12}}
=\sum_{all\ \lambda}
A_{\lambda_1,\lambda_2,\lambda_{12}}(r_1,r_2,R) 
\,Y_{\lambda_1,\lambda_2,\lambda_{12}}(\hat{r}_1,\hat{r}_2,\hat{R})
\end{equation}
with 
\begin{equation}
Y_{\lambda_1,\lambda_2,\lambda_{12}}(\hat{r}_1,\hat{r}_2,\hat{R})
%=\sum_{m_{\lambda_1},m_{\lambda_2},m_{\lambda_{12}}}
=\sum_{all\ m}
\langle \lambda_1 m_{\lambda_1} \lambda_2 m_{\lambda_2}|
\lambda_{12}m_{\lambda_{12}}\rangle\,
Y_{\lambda_1m_{\lambda_1}}(\hat{r}_1)\,
Y_{\lambda_2m_{\lambda_2}}(\hat{r}_2)\,
Y^*_{\lambda_{12}m_{\lambda_{12}}}(\hat{R}) ,
\end{equation}
where $\langle ...|...\rangle$ represents a Clebsch-Gordan coefficient
and $Y_{\lambda m}(\hat{r})$ is a spherical harmonic.
The total wave function for the four-atom system is 
expanded in terms of a diabatic basis set which
contains products of molecular wave functions $\chi_{v_ij_i}(r_i)$
with vibrational and rotational quantum numbers 
$v_i$ and $j_i$, respectively. We describe the
combined molecular state (CMS) comprised of 
CO($v_1,j_1$) and H$_2(v_2,j_2)$ using the
notation $n=(v_1,j_1,v_2,j_2)$ so that the 
basis functions for vibrational motion may be written
\begin{equation}
\chi_n(r_1,r_2)=\chi_{v_1j_1}(r_1)\chi_{v_2j_2}(r_2)\ .
\end{equation}
The rotational wave functions are given in terms of products
of spherical harmonics in a total angular momentum representation.
The basis sets are defined by the number of vibrational levels
and the maximum rotational level $j_v^{max}$ for a given $v$.
The radial interaction potential matrix elements
are obtained by integrating over the internal coordinates
\begin{eqnarray}
B^{\lambda_1,\lambda_2,\lambda_{12}}_{n;n'}(R)=
\int_0^{\infty}\int_0^{\infty}\chi_{n}(r_1,r_2)
A_{\lambda_1,\lambda_2,\lambda_{12}}(r_1,r_2,R)
\chi_{n'}(r_1,r_2)r_1^2r_2^2dr_1dr_2\ .
\end{eqnarray}
The full potential matrix depends on the 
scattering formulation. For the CC method, the
channels are defined by the set $\{n,j_{12},l\}$,
where $l$ is the orbital angular momentum quantum number
and $\vec{j}_{12}=\vec{j}_1+\vec{j}_2$. 
The total angular momentum quantum number $J$ is defined
by $\vec{J}=\vec{l}+\vec{j}_{12}$,
and the interaction potential matrix for 6D-dynamics is given by
\newpage
\begin{eqnarray}
&&U^{J}_{nj_{12}l;n'j_{12}'l'}(R)=(4\pi)^{-3/2}
%\sum_{\lambda_1,\lambda_2,\lambda_{12}}(-1)^{j'_1+j'_2+j_{12}+J}
\sum_{all\ \lambda}(-1)^{j'_1+j'_2+j_{12}+J}
\left([j_1][j_2][j_{12}][l][j'_1][j'_2][j'_{12}][l'][\lambda_1][\lambda_2]
[\lambda_{12}]^2\right)^{1/2}
\nonumber\\ 
& \times &
\left(\begin{array}{ccc} l' & \lambda_{12} & l\\ 0 & 0 & 0 \\ \end{array} \right)
\left(\begin{array}{ccc} j'_1 & \lambda_1 & j_1\\ 0 & 0 & 0 \\ \end{array} \right)
\left(\begin{array}{ccc} j'_2 & \lambda_2 & j_2\\ 0 & 0 & 0 \\ \end{array} \right)
\left\{\begin{array}{ccc} l & l' & \lambda_{12} \\ j'_{12} & j_{12} & J \\ 
\end{array}\right\}
\left\{\begin{array}{ccc} j_{12} & j_2 & j_1 \\ j'_{12} & j'_2 & j'_1 \\ 
\lambda_{12} & \lambda_2 & \lambda_1 \\ \end{array} \right\}
B^{\lambda_1,\lambda_2,\lambda_{12}}_{n;n'}(R)
%\nonumber\\
\label{V6}
\end{eqnarray}
which is diagonal with respect to $J$ and independent of 
the projection of $\vec{J}$ in the space-fixed frame.
The notations $(===)$, $\{===\}$, and $\{\equiv\equiv\equiv\}$
are the usual 3j, 6j, and 9j symbols, and $[j]=(2j+1)$.
For the 5D-CS formulation, the potential matrix is given by
\begin{eqnarray}
&&U^{m_1,m_2}_{n;n'}(R)=(4\pi)^{-3/2}
%\sum_{\lambda_1,\lambda_2,\lambda_{12}}(-1)^{\lambda_1+\lambda_2+m_1+m_2}
\sum_{all\ \lambda}(-1)^{\lambda_1+\lambda_2+m_1+m_2}
\left([j_1][j_2][j'_1][j'_2][\lambda_1][\lambda_2][\lambda_{12}]^2\right)^{1/2}
\nonumber\\ & \times &
\left(\begin{array}{ccc} j'_1 & \lambda_1 & j_1\\ 0 & 0 & 0 \\ \end{array} \right)
\left(\begin{array}{ccc} j'_2 & \lambda_2 & j_2\\ 0 & 0 & 0 \\ \end{array} \right)
\left(\begin{array}{ccc} j'_1 & \lambda_1 & j_1\\ -m_1 & 0 & m_1 \\ \end{array} \right)
\left(\begin{array}{ccc} j'_2 & \lambda_2 & j_2\\ -m_2 & 0 & m_2 \\ \end{array} \right)
\left(\begin{array}{ccc} \lambda_1 & \lambda_2 & \lambda_{12}\\ 0 & 0 & 0 \\ \end{array} \right)
B^{\lambda_1,\lambda_2,\lambda_{12}}_{n;n'}(R)\nonumber\\
\label{V5}
\end{eqnarray}
which is diagonal with respect to $m_1$ and $m_2$, the projection quantum numbers 
of $\vec{j}_1$ and $\vec{j}_2$, respectively. The 5D-CS formulation averages the
twist-angle dependence of the PES and yields a potential matrix which is independent of $j_{12}$.
Both dynamical formulations require the solution of a set of coupled radial equations
derived from the Schr\"odinger equation.
Cross sections 
%at a given collision energy $E_c$
may be expressed in terms of the appropriate $T$-matrix by
\begin{eqnarray}
\sigma_{n\rightarrow n'}^{\tiny{\mbox{6D-CC}}}(E_c)
=\frac{\pi}{(2j_1+1)(2j_2+1)2\mu E_c} 
\sum_{j_{12}j'_{12}ll'J} (2J+1)\,
\left|T^{J}_{nj_{12}l;n'j_{12}'l'} (E_c)\right|^{2}\ ,
\label{S6}
\end{eqnarray}
and 
\begin{eqnarray}
\sigma_{n\rightarrow n'}^{\tiny{\mbox{5D-CS}}}(E_c)
=\frac{\pi}{(2j_1+1)(2j_2+1)2\mu E_c} 
\sum_{\bar{l}m_1m_2} (2\bar{l}+1)\,
\left|T^{\bar{l}m_1m_2}_{n;n'} (E_c)\right|^{2}\ ,
\label{S5}
\end{eqnarray}
where $E_c$ is the collision energy, and $\mu$ is the reduced mass of the CO-H$_2$ system.
%where $\mu$ is the reduced mass of the CO-H$_2$ system.
The CS approximation assumes the off-diagonal Coriolis matrix elements 
of $\hat{l}^2$ with respect to $m_1$ and $m_2$ may be neglected, and the diagonal 
elements may be approximated by an effective orbital angular momentum quantum number 
$\bar{l}$ which replaces $l$.  For all calculations in the present work, 
$\overline{l}\equiv J$ which is its average value between $|J-j_{12}|$ and $J+j_{12}$.
Rate coefficients 
%at a temperature $T$
may be obtained by thermally averaging the cross sections over a Maxwellian
velocity distribution
\begin{eqnarray}
k_{n\rightarrow n'}(T)=\sqrt{\frac{8k_BT}{\pi\mu}}\left(k_BT\right)^{-2}
\int_0^\infty \sigma_{n\rightarrow n'}(E_c)\,e^{-E_c/k_BT}E_c\, dE_c\ ,
\label{rate}
\end{eqnarray}
where $T$ is the temperature and $k_B$ is Boltzmann's constant.

\section{Results}

Previous work \cite{forrey} showed that 5D-CS and 6D-CS results for CO+H$_2$
were virtually identical for collision energies above 10 cm$^{-1}$.
The agreement with 6D-CC calculations was also found to be good at these energies
for the limited amount of data that was available for comparison.
Recently, new 6D-CC results have been reported \cite{yang} 
which allow for more extensive testing of the 5D-CS approximation. 
All calculations in the present work were performed using a modified 
version of the TwoBC code \cite{krems} which replaces equations 
(\ref{V6}) and (\ref{S6}) with equations (\ref{V5}) and (\ref{S5}).
The radial coordinates of both molecules were represented as
discrete variables with 20 points each. Gauss-Legendre quadratures
were used for $\theta_1$ and $\theta_2$ with 14 points each, and
a Chebyshev quadrature was used for $\phi$ with 8 points.
The maximum values for $\lambda_1$ and $\lambda_2$ were 10 and 6,
respectively.  The log-derivative matrix propagation method 
\cite{johnson,mano} was used to integrate the set of coupled
equations derived from the Schr\"odinger equation from $R=4-18$ a.u. 
in steps of 0.05 a.u. 
The calculations were performed in unit steps for four    
decades of collision energy on a logarithmic energy grid.
The maximum effective orbital angular momentum for each set
of calculations is given by
\begin{equation}
\overline{l}_{max}=\left\{\begin{array}{l}
20, \ \ \ E_c=1-10 \ \mbox{cm}^{-1}\ ,\\
40, \ \ \ E_c=10-100 \ \mbox{cm}^{-1}\ ,\\
80, \ \ \ E_c=100-1,000 \ \mbox{cm}^{-1}\ ,\\
160, \ \ E_c=1,000-10,000 \ \mbox{cm}^{-1}\ .
\end{array}\right.
\end{equation}
Excellent agreement was seen at the $E_c$ boundaries, 
and convergence tests at low $E_c$ verified that the results 
were converged to within 5\% or better with respect to size of the basis set. 

The 5D-CS calculations were performed for collision energies
between 10-10,000 cm$^{-1}$ and compared against 6D-CC results
obtained perviously \cite{yang}. Figure 2 shows elastic and 
rotationally inelastic cross sections for the $(1,1,0,0)$ initial
state. The curves for rotational excitation $(1,2,0,0)$ and 
de-excitation $(1,0,0,0)$ are in near perfect agreement with 
the corresponding 6D-CC points for energies above 100 cm$^{-1}$. 
The agreement remains good for energies down to 10 cm$^{-1}$, however, 
the 5D-CS results do not show the resonant structure that was found 
in the 6D-CC calculations. This is partly due to the energy step
size, which is too large to resolve the resonances, however, it
is not expected that the effective $\overline{l}$ used in the
CS calculations would accurately describe the resonances even
if a finer energy grid were used.
The elastic $(1,1,0,0)$ cross section
shows a similar pattern of agreement with the exception of small
discrepancies at 500 and 1000 cm$^{-1}$. This is probably due to
incomplete convergence of the 6D-CC elastic cross section with 
respect to higher partial waves. 
%Elastic cross sections are generally slower to converge 
%and are not normally needed in astrophysical models.

Figure 3 shows a set of similar comparisons for the 
$(1,2,0,0)$, $(1,3,0,0)$, $(1,4,0,0)$, and $(1,5,0,0)$
initial states. Only the rotational de-excitation cross sections 
are shown. The agreement between the 5D-CS and 6D-CC results
is again very good for energies between 10-1000 cm$^{-1}$.
Apart from a few small discrepancies, the approximate 5D-CS
formulation is able to reproduce the results of the 6D-CC 
calculation and extend them to higher energies.
The agreement is not quite as good for the vibrationally
inelastic cross sections shown in Figure 4. Cross sections
for transitions from the same initial states as Figures 2 and 3 
are shown for selected final states. The total cross section
represents the sum over all final CO rotational level 
contributions in the $v_1=0$ manifold. The state-to-state
cross sections show a larger number of discrepancies
between the 5D-CS and 6D-CC calculations. However, 
these discrepancies are generally small, and the
5D-CS results are able to provide a good approximation
to the overall shape and relative magnitudes of the 
state-to-state cross sections for vibrationally
inelastic collisions. The agreement between the two
computational methods for the total cross section
appears to worsen at high energies as the initial 
rotational level of the CO molecule increases. 
This is due to the contributions from transitions 
to high rotational levels. For example, the
$(1,5,0,0)$ to $(0,10,0,0)$ cross section in Figure 4(f) 
shows a relatively large discrepancy compared to the 
$(0,0,0,0)$ and $(0,2,0,0)$ cross sections which are
in very good agreement at high energies. This pattern
was found for all of the vibrationally inelastic
cross sections and may be attributed to basis set
truncation error in the 6D-CC calculations which
used $j_{v_1=0}^{max}=22$ and $j_{v_1=1}^{max}=20$.
These limits provide good convergence
for vibrationally inelastic cross sections involving
low-lying rotational levels at all energies. However,
for transitions involving high rotational levels, the
truncation limits are insufficient, particularly at
high collision energies.
It should be noted that the truncation in the 6D-CC
calculations was a necessity due to the computational
resource limitations.
For comparison, the 5D-CS calculations in Figures 2-4
used $j_{v_1=0}^{max}=j_{v_1=1}^{max}=40$.  

The size of the CO vibrational basis set is considered
in Figure 5. For both the 5D-CS and 6D-CC calculations,
the vibrational basis set included $v_1=0-2$. 
The vibrational quenching cross section for the $(2,0,0,0)$ 
initial state is dominated by the $\Delta v_1=-1$ contribution,
and the comparison between the 5D-CS and 6D-CC results is
similar to the $(1,0,0,0)$ initial state shown in Figure 4.
The $\Delta v_1=-2$ contribution is at least two orders of
magnitude smaller than the $\Delta v_1=-1$ contribution
for both sets of calculations at the energies shown. 
The agreement between the 5D-CS and 6D-CC results is 
not as good for the $\Delta v_1=-2$ contribution 
above 300 cm$^{-1}$, but given the relatively small 
magnitude of this cross section, the discrepancy is
not very significant.  Expanding the basis set to include 
closed vibrational levels yielded no change to these results. 
Figure 5b shows similar results for the $(5,0,0,0)$ initial
state. These cross sections are again
dominated by the $\Delta v_1=-1$ contribution, however, the
$\Delta v_1=-2$ contribution increases relative to the
$\Delta v_1=-1$ contribution due to the decreasing difference 
between vibrational energy levels.
The small contribution from $\Delta v_1=-2$ transitions allows
a compact representation of the vibrational motion
to be obtained using basis sets with $v_1$ equal to $v$ 
and $v-1$, where $v$ is the vibrational level of the initial 
state. This enables large rotational basis sets 
to be employed (see Table I). 

The size of the para-H$_2$ rotational basis set is considered 
in Figure 6. The solid curves correspond to a rotational
basis set which includes $j_2=0-4$.
In order to improve the computational efficiency
of these calculations, the truncation limit $j_{v_1=1}^{max}$
of the initial CO rotational basis was reduced to 20.
The figure shows that the cross sections for vibrational 
relaxation of CO accompanied by rotational excitation of H$_2$
are comparable in magnitude to those with no change in H$_2$.
Expanding the H$_2$ basis further to allow more rotational
excitation yielded small contributions and provided
little change to these results. The points shown in
the figure correspond to the $j_{v_1}^{max}=40$ basis set
with H$_2$ restricted to its ground rovibrational state.
The agreement in the H$_2(0,0)$ cross sections is nearly
perfect for the two sets of calculations for energies
below 2000 cm$^{-1}$. At higher energies, it is not
clear which basis set yields the more reliable result
due to the different truncation limits on H$_2$ and CO. 

In order to explore rovibrational transitions for a
large sample of internal states and collision energies,
we constructed two separate databases for state-to-state 
cross sections and rate coefficients.  The ``frozen H$_2$" 
database uses $j_{v}^{max}=40$ and includes transitions 
for initial CO$(j_1=0-30)$
due to collisions with rovibrationally rigid para-H$_2(j_2=0)$ 
and ortho-H$_2(j_2=1)$.
%It also includes $j_{0}^{max}=40$ for all of the $v=2$ 
%calculations.
The ``flexible H$_2$" database uses $j_{v}^{max}=20$ 
and includes transitions for initial
CO$(j_1=0-10)$ due to collisions with rotationally
flexible para-H$_2(j_2=0,2,4)$ and ortho-H$_2(j_2=1,3,5)$.
Additional details of the basis sets are given in Table I.

Figures 7-9 show rate coefficients for a sample of
initial states from the frozen-H$_2$ database.
The pure rotational relaxation rate coefficients in Figure 7
are very similar to the rigid rotor results reported 
previously \cite{rotor}. The present results are for $v=1$, 
however, similar plots for higher $v$ are nearly identical,
which confirms the validity of the rigid rotor approximation
for pure rotational transitions of CO. For the rovibrationally
inelastic transitions shown in Figures 8 and 9, the curves are
well-ordered with respect to the initial vibrational level and
generally exhibit a step-like structure with increasing temperature.
The height of the steps increases with final state rotational level
$j_1'$, and the rate coefficients are found to be relatively large for
$j_1'$ near the basis set truncation limit at the highest temperatures 
shown. This indicates that the rovibrationally inelastic calculations 
are still not fully converged at the highest temperatures ($T>1000$ K)
even with the large truncation limit $j_{v-1}^{max}=40$.

Figure 10 shows results from the flexible-H$_2$ database for the
$(v,10,0,0)$ initial state. Rate coefficients for transitions 
involving no change in H$_2$ are nearly identical to those 
presented in Figure 8 and are not shown. Vibrational relaxation
of CO$(v=1-5)$ for transitions which leave H$_2$ in a rotationally 
excited state yield rate coefficients which are well-ordered with
respect to $v$ as shown. The solid black and dashed red curves 
represent $\Delta j_2=2$ and $\Delta j_2=4$ transitions for H$_2$, 
respectively, with both sets of curves increasing with $v$.
The step-like structure is again seen in the rate coefficients
and is a general feature for both databases.

The databases include CO collisions with both para-H$_2$ and ortho-H$_2$.
%Figure 11 shows the ortho/para ratio of total vibrational quenching
Figure 11 shows the ratio of total CO vibrational quenching
rate coefficient for ortho-H$_2$ and para-H$_2$ collisions
%versus temperature for several initial states of CO. 
for several initial states of CO. 
The ratio is generally found to approach unity for $T>100$ K. 
As the initial rotational level of CO is increased, the ratio
is close to unity for all vibrational levels and temperatures
shown. These results suggest that para-H$_2$
rate coefficients may be directly used for ortho-H$_2$ rate
coefficients for $T>100$ K with an error of no more than 10\% 
and a considerable savings in computational expense.

\section{Conclusions}

Full-dimensional dynamics of CO interacting with H$_2$ 
is a challenging computational problem for quantum mechanical 
scattering formulations. The closely spaced rotational levels 
for CO lead to substantial internal angular momentum coupling 
between the molecules. When orbital angular momentum is properly
taken into account within the numerically exact CC formulation,
the additional angular momentum coupling creates a bottleneck
for many rovibrationally excited initial states, which makes 
it impossible to bring the calculations to completion on a 
practical timescale. In such cases, it is necessary to use
approximate methods to obtain the desired data.
The CS approximation offers a good compromise between accuracy 
and computational effort. In this approximation, the exact 
orbital angular momentum quantum number is replaced by an 
effective value, and the off-diagonal Coriolis coupling 
between states with different projection quantum numbers 
is neglected. Averaging the PES over the twist angle to
obtain the 5D-CS approximation further improves the
computational efficiency without introducing a 
significant additional loss of accuracy beyond the
CS approximation itself.

Using the 5D-CS approximation, we constructed two separate
databases for state-to-state cross sections and rate coefficients
for CO+H$_2$ collisions. The ``frozen-H$_2$" database assumes the
collision takes place with the H$_2$ molecule frozen in the lowest
rotational level of a given symmetry. This restriction allows the
bulk of the computational effort to be used for the CO basis sets
which include $j_v^{max}=j_{v-1}^{max}=40$ for each initial $v$. 
The database then includes transitions for each
basis set with initial CO$(v=0-5, j=0-30)$. The ``flexible-H$_2$"
database allows rotational excitation of H$_2$ during the collision. 
To compensate for the enlarged H$_2$ basis sets, the CO basis sets 
are reduced to $j_v^{max}=20$ with $j_{v-1}^{max}=40$, and a smaller
set of transitions from initial CO$(v=0-5, j=0-10)$ is included
in the database. The two databases are in good agreement with
each other for transitions that are common to both basis sets.
Both databases generally contain cross sections 
and rate coefficients for 
$E_c=1-10,000$ cm$^{-1}$ and $T=10-3,000$ K.
%with the understanding that the rate coefficients are
%fully converged with respect to the Maxwellian distribution
%only for $60\le T\le 3000$ K. Temperatures outside this range
%are included as a starting point for further development of
%the databases.  A numerical evaluation of 
%equation (\ref{rate}) with the product of cross section and velocity
%set to unity is provided for each temperature in order to determine
%whether the energy range is sufficient to obtain proper
%normalization. Generally, the normalization computed in 
%this way will be smaller than unity 
%for $T<60$ K and $T>3000$ K 
%unless additional energies are included in the integration. 
%It would be desirable to improve the databases by adding 
%6D-CC cross sections for the lower energies, and by 
It would be desirable to improve the databases by replacing
the 5D-CS cross sections with 6D-CC values for the lower energies, 
and by expanding the 5D-CS basis sets to include additional 
CO rotational levels at the higher energies.
These extensions will be implemented in future versions of the
project. The present version provides a large amount of data
which may be used as a starting point for further 
approximations (e.g. scaling) and provides
important data for astrophysical models.
The databases may be downloaded from the
supplementary material associated with this article
and from the UGA Excitation Database\cite{UGA} 
or PSU website\cite{PSU}.

\begin{acknowledgments}
This work was supported at Penn State
by NSF Grant No. PHY-1503615, at
UGA by NASA grant NNX12AF42G, and
at UNLV by NSF Grant No. PHY-1505557.
\end{acknowledgments}

%\newpage
\vspace*{.5in}

\begin{table}[h]
\centering
\caption{Basis set parameters used for the databases.}
\vspace*{.2in}
\begin{tabular}{|c|c|c|c|c|c|c|c|c|}\hline
basis set & $v$ & \ $j_v^{max}$ \ & \ $j_{v-1}^{max}$ \ & \ $j_{v-2}^{max}$ \ & 
$j_1$ & $j_2$ & $j_1'$ & $j_2'$ \ \\ \cline{1-9}
frozen-H$_2$ & 0 & 40 & - & - & 0-30 & 0, 1 & 0-40 & 0, 1 \\ 
             & 1 & 40 & 40 & - & 0-30 & 0, 1 & 0-40 & 0, 1 \\ 
             & 2 & 40 & 40 & 40 & 0-20 & 0, 1 & 0-40 & 0, 1 \\ 
             & 3-5 & 40 & 40 & - & 0-20 & 0, 1 & 0-40 & 0, 1 \\ \cline{1-9}
$\ $flexible-H$_2\ $ & 0 & 20 & - & - & 0-10 & 0, 1 & 0-20 & 0-4, 1-5 \\ 
               &\ 1-2 \ & 20 & 40 & - & \ 0-10 \ &  \ 0, 1 \ & \ 0-40 \ & \ 0-4, 1-5 \ \\ 
               &\ 3-5 \ & 20 & 40 & - & \ 0-10 \ &  \ 0 \ & \ 0-40 \ & \ 0-4 \ \\ \cline{1-9}
\end{tabular}
\end{table}

\newpage

\begin{figure}
\centerline{\epsfxsize=6in\epsfbox{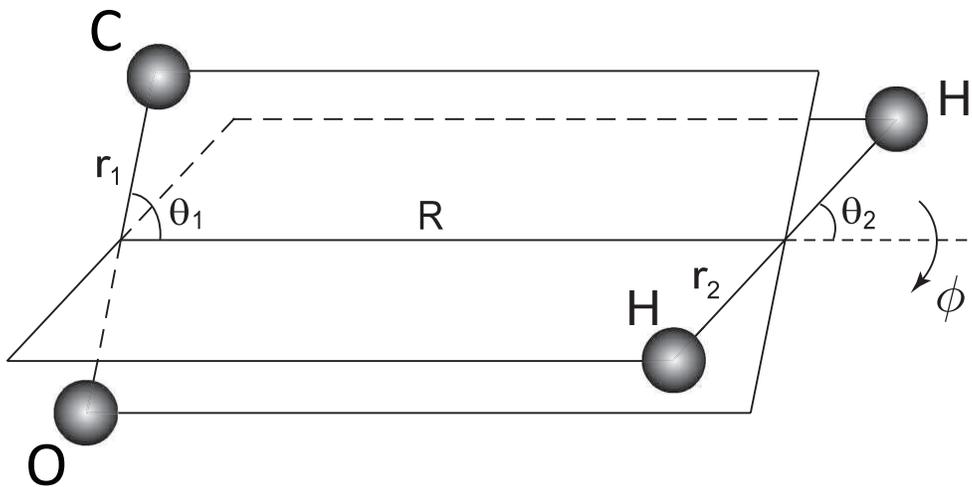}}
\caption{Six-dimensional Jacobi coordinates for CO-H$_2$. 
$R$ is the distance between the centers of mass of CO and H$_2$, 
$r_1$ and $r_2$ are bond lengths, $\theta_1$ and $\theta_2$ 
are the respective angles between $\vec{R}$ and $\vec{r}_1$ 
and $\vec{r}_2$, and $\phi$ is the dihedral or twist angle.}
\end{figure}

\newpage

\begin{figure}
\centerline{\epsfxsize=6in\epsfbox{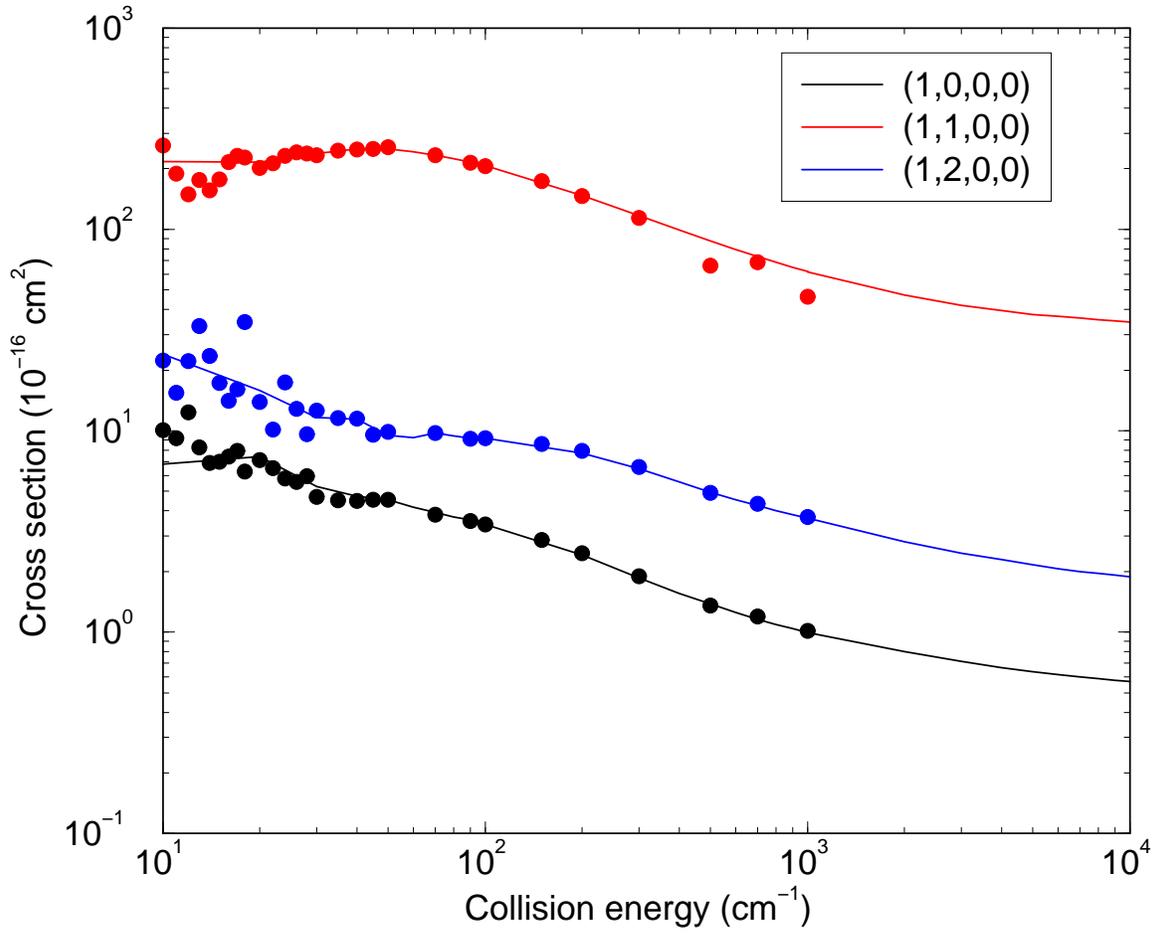}}
\caption{Elastic and rotationally inelastic cross sections
for inital state $(1,1,0,0)$.
Solid lines are 5D-CS results and points are 6D-CC results \cite{yang}.}
\end{figure}

\newpage

\begin{figure}
\centerline{\epsfxsize=3in\epsfbox{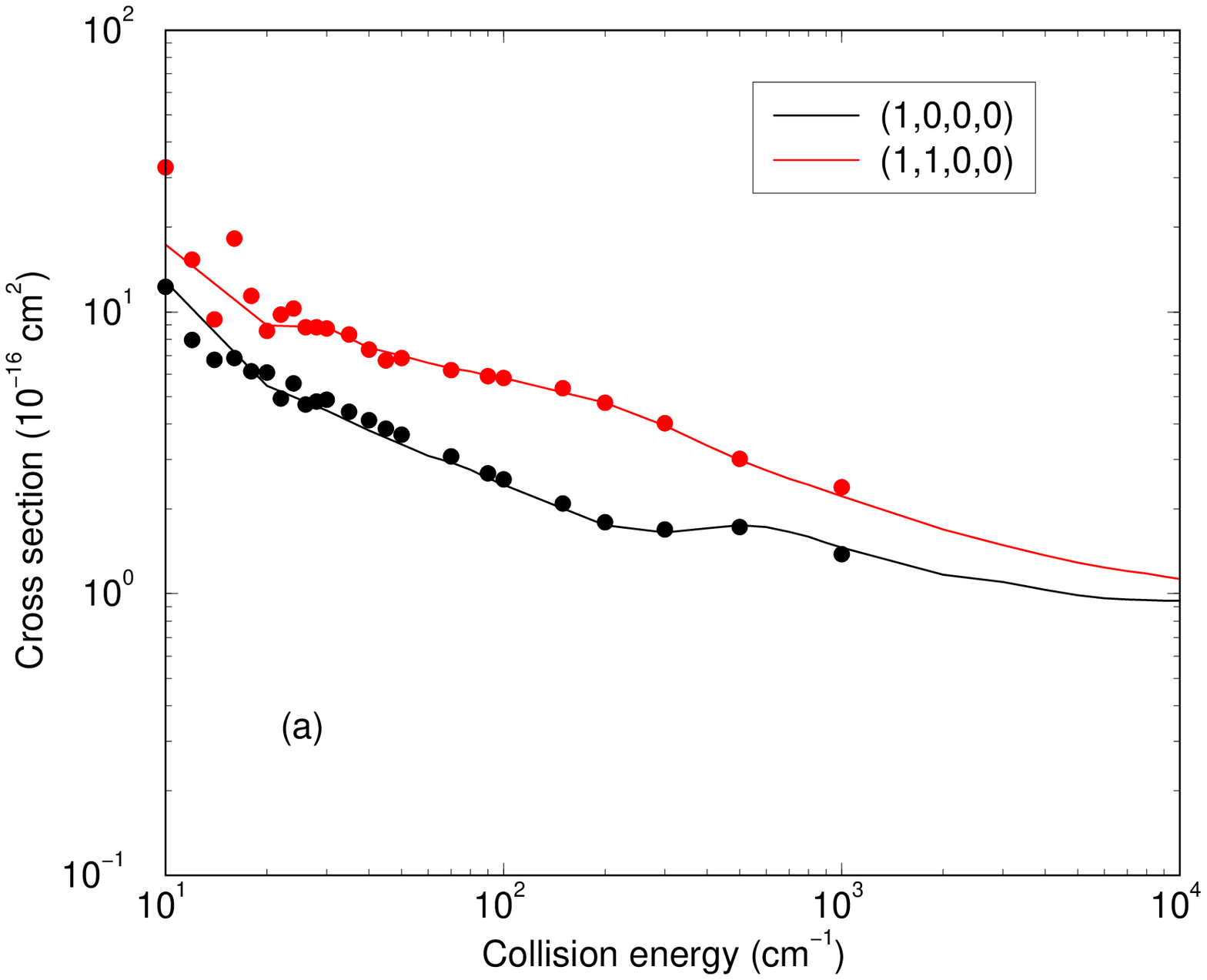}\hspace{.1in}\epsfxsize=3in\epsfbox{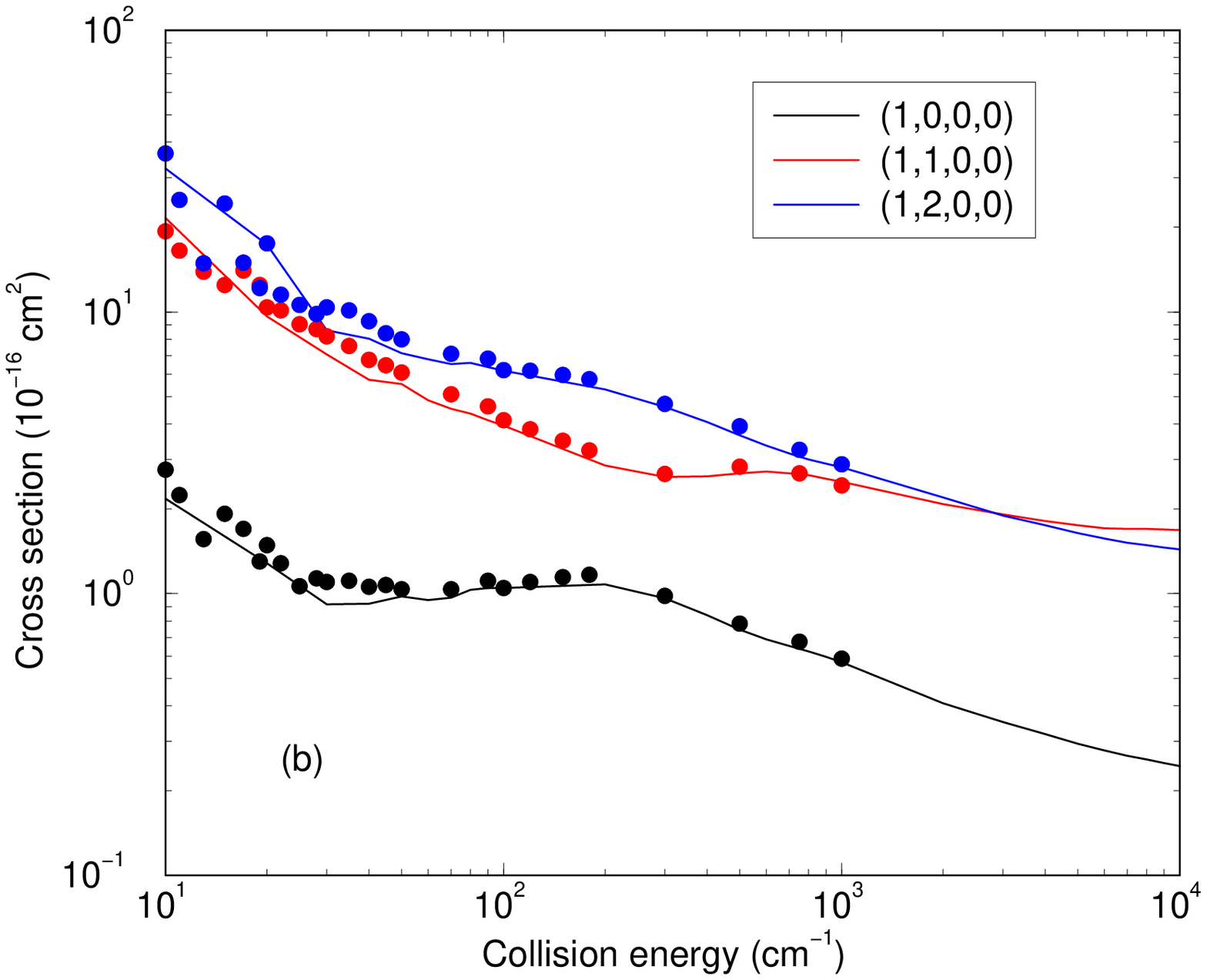}}
\centerline{\epsfxsize=3in\epsfbox{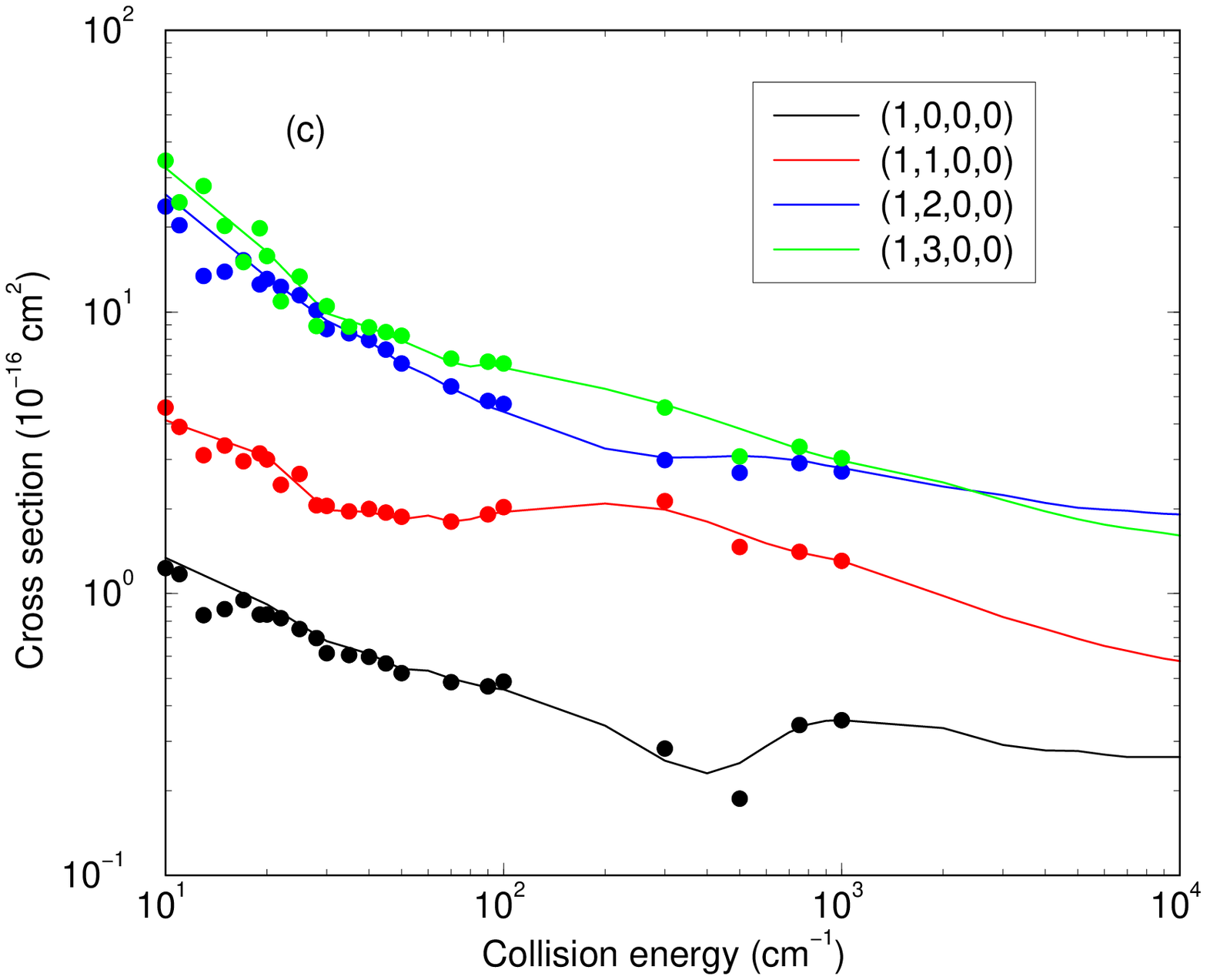}\hspace{.1in}\epsfxsize=3in\epsfbox{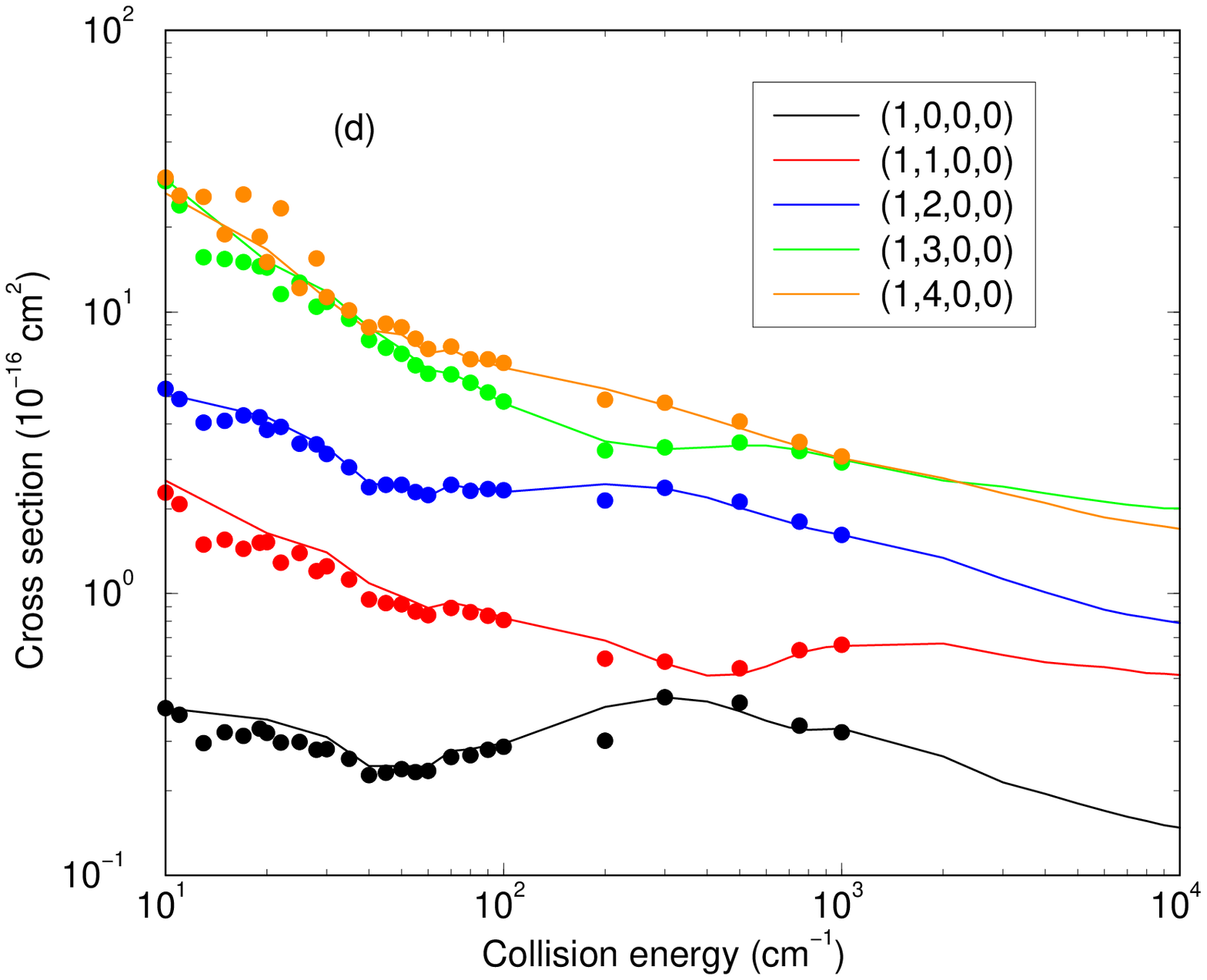}}
\caption{Rotationally inelastic cross sections for transitions 
between the initial state 
(a) $(1,2,0,0)$, 
(b) $(1,3,0,0)$, 
(c) $(1,4,0,0)$,
(d) $(1,5,0,0)$
and the final state indicated. 
Solid lines are 5D-CS results and points are 6D-CC results \cite{yang}.}
\end{figure}

\newpage

\begin{figure}
\centerline{\epsfxsize=3in\epsfbox{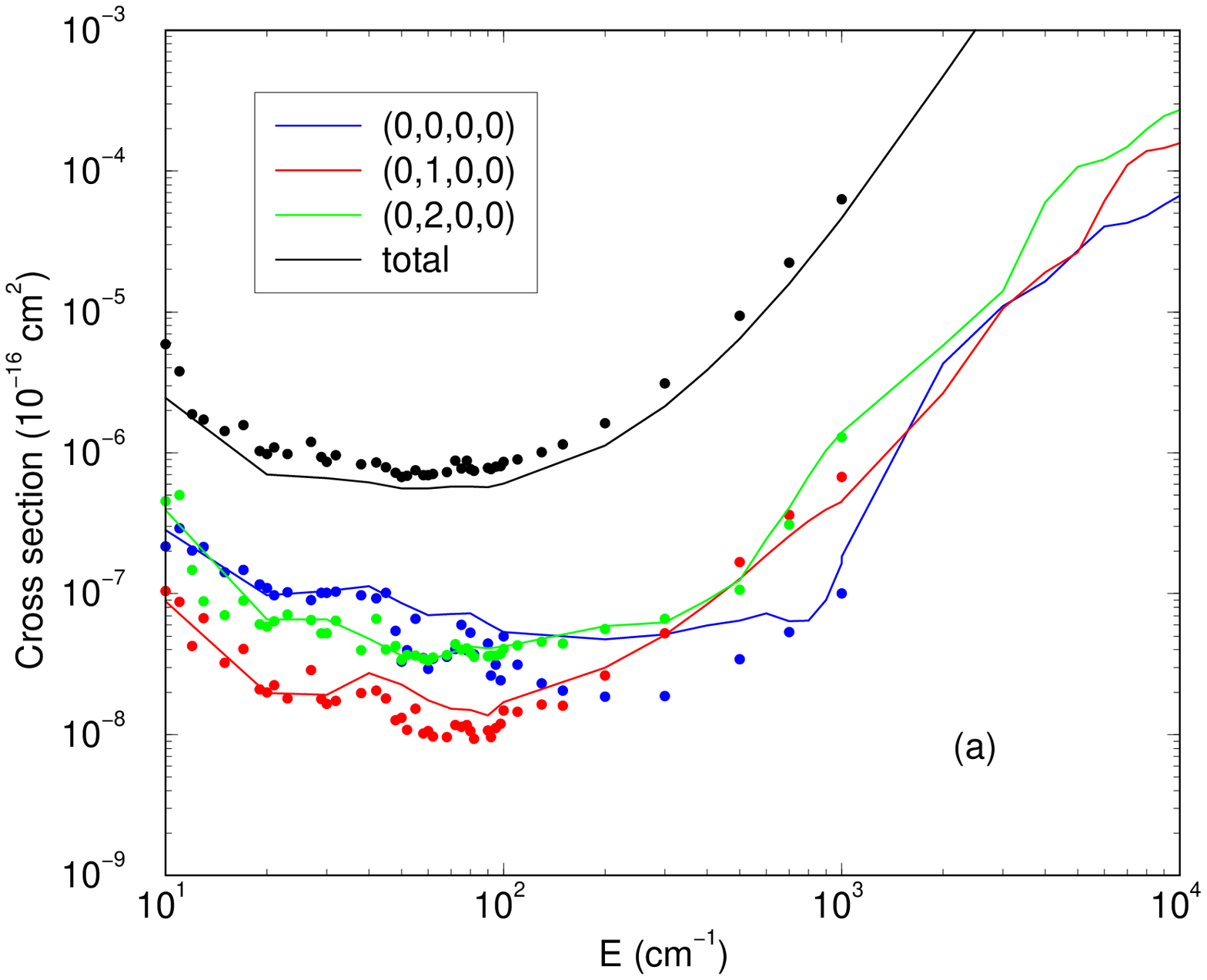}\hspace{.1in}\epsfxsize=3in\epsfbox{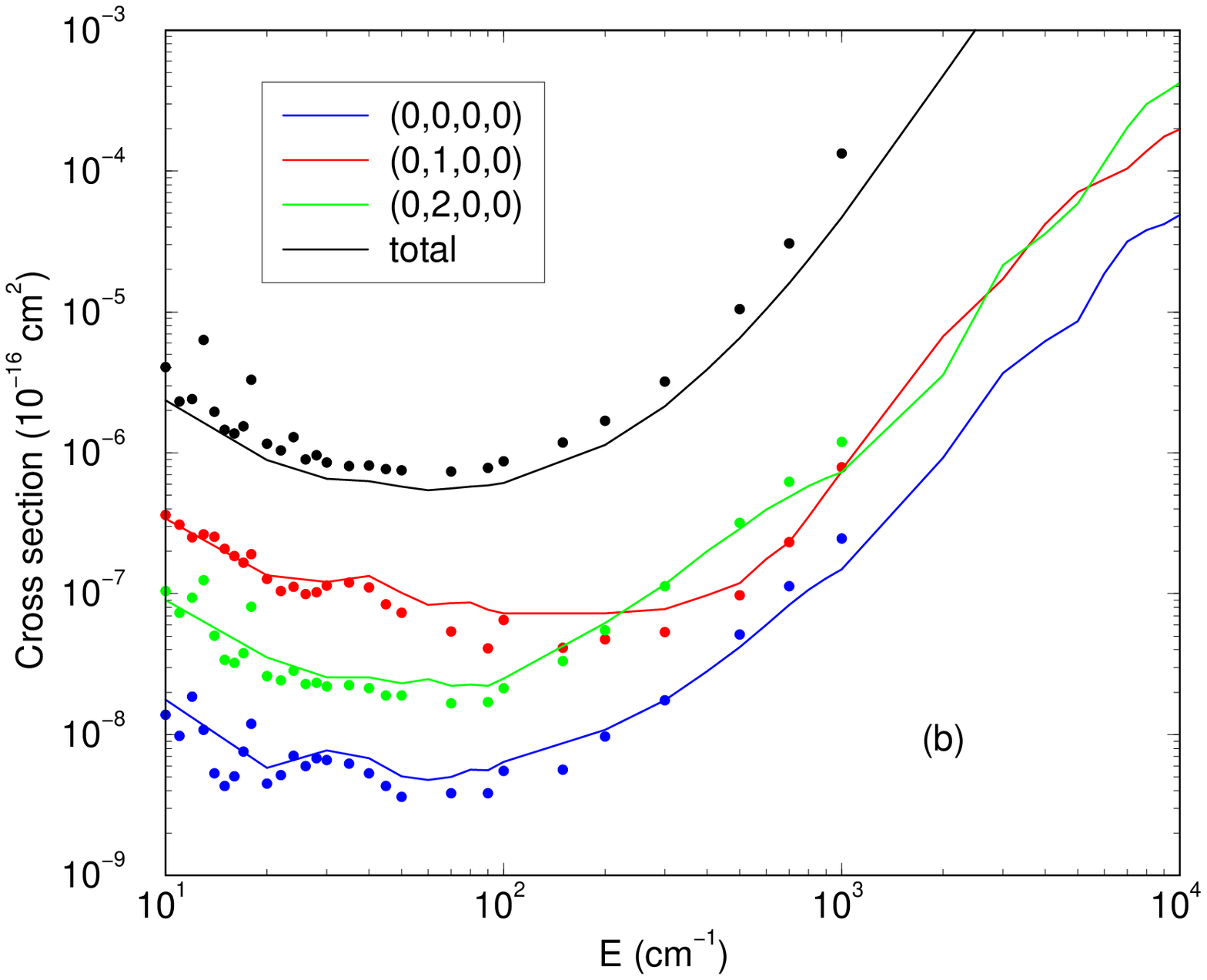}}
\centerline{\epsfxsize=3in\epsfbox{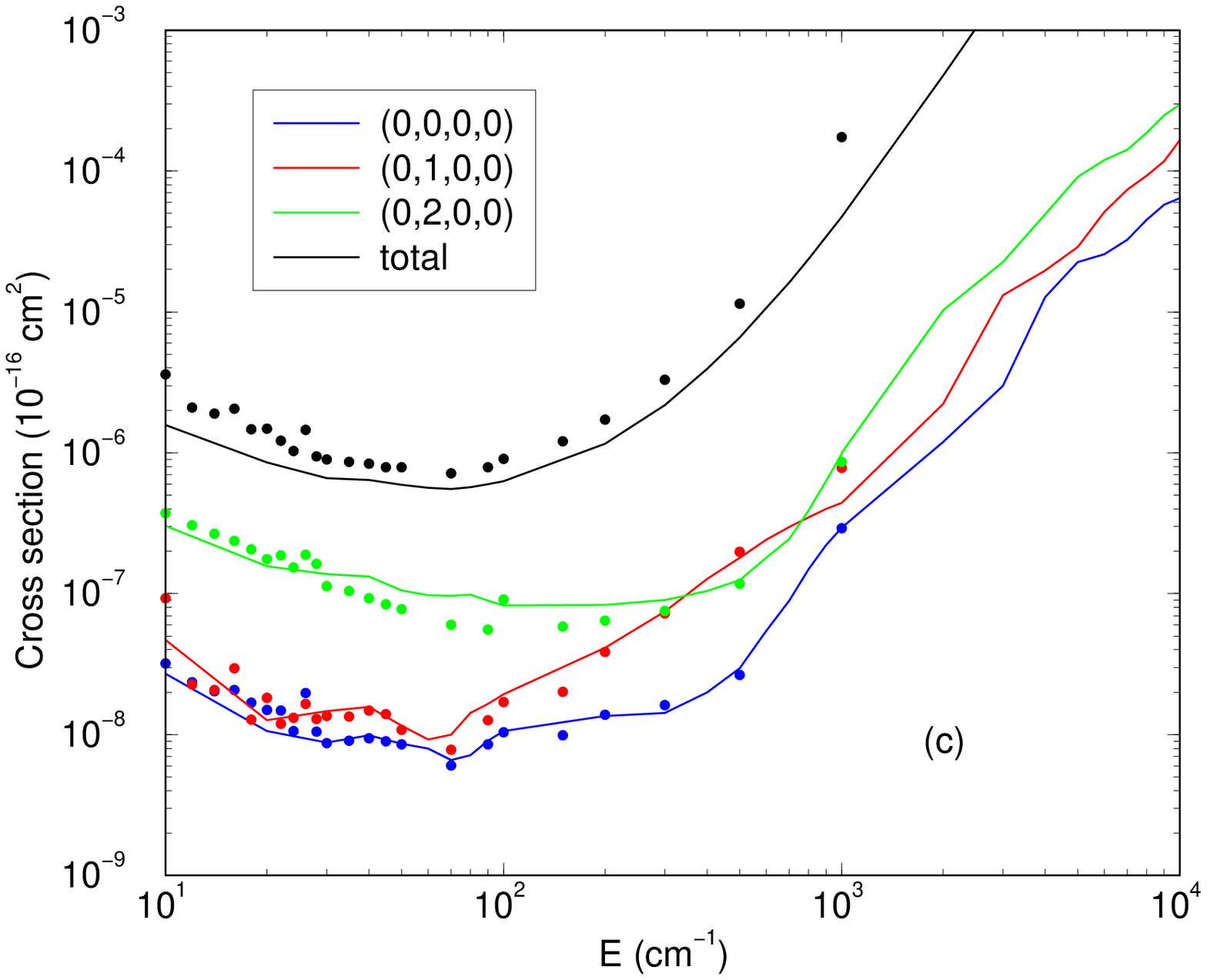}\hspace{.1in}\epsfxsize=3in\epsfbox{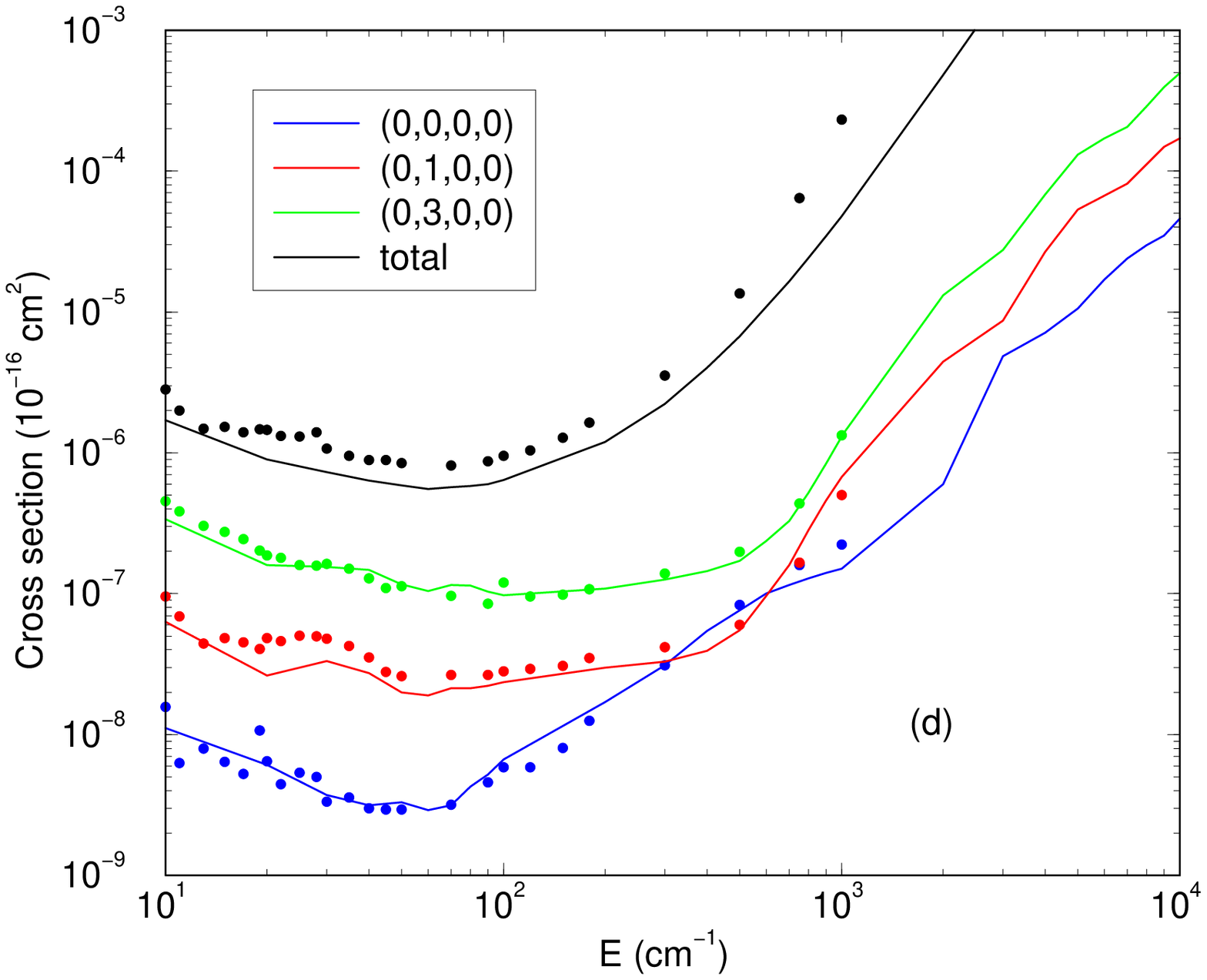}}
\centerline{\epsfxsize=3in\epsfbox{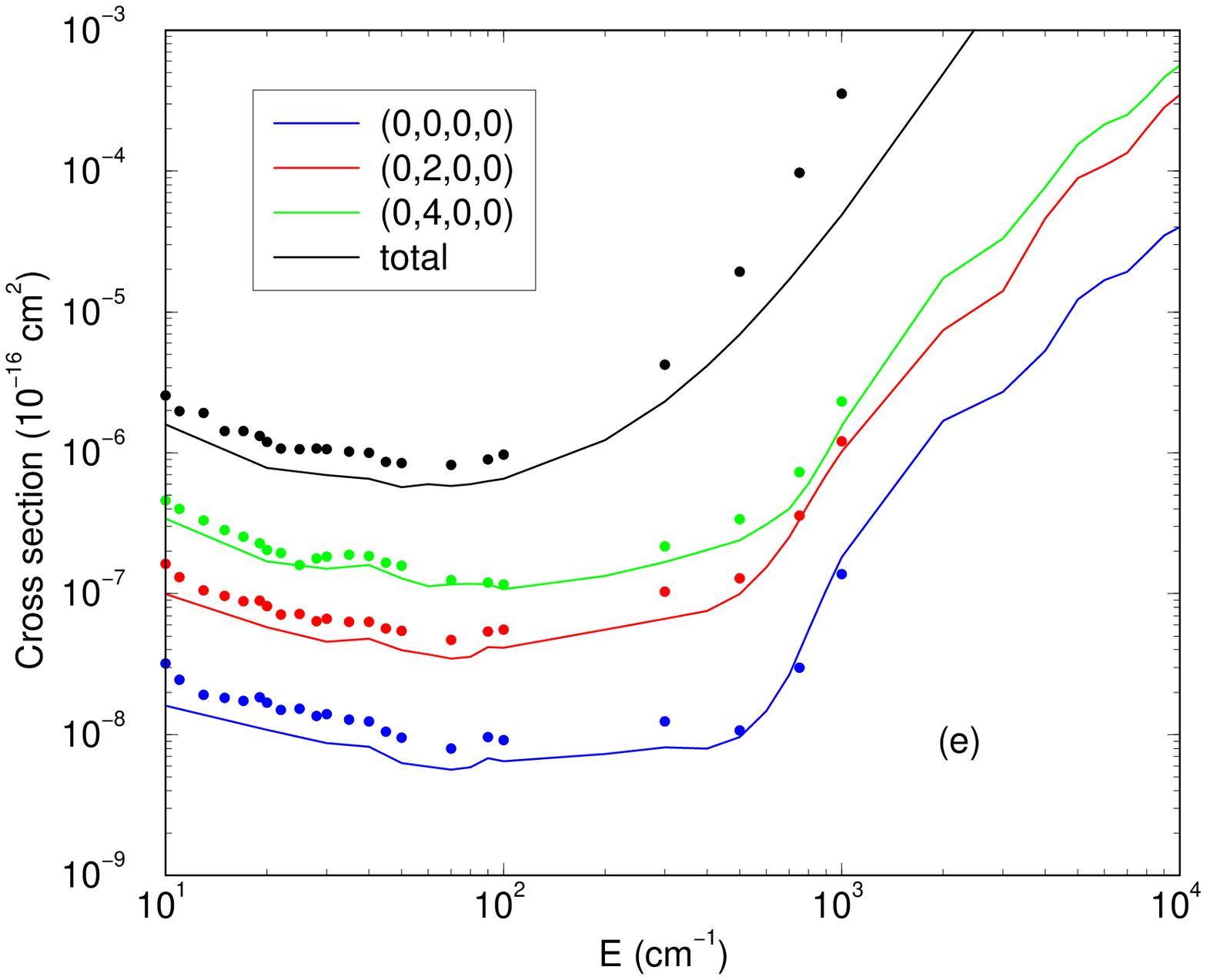}\hspace{.1in}\epsfxsize=3in\epsfbox{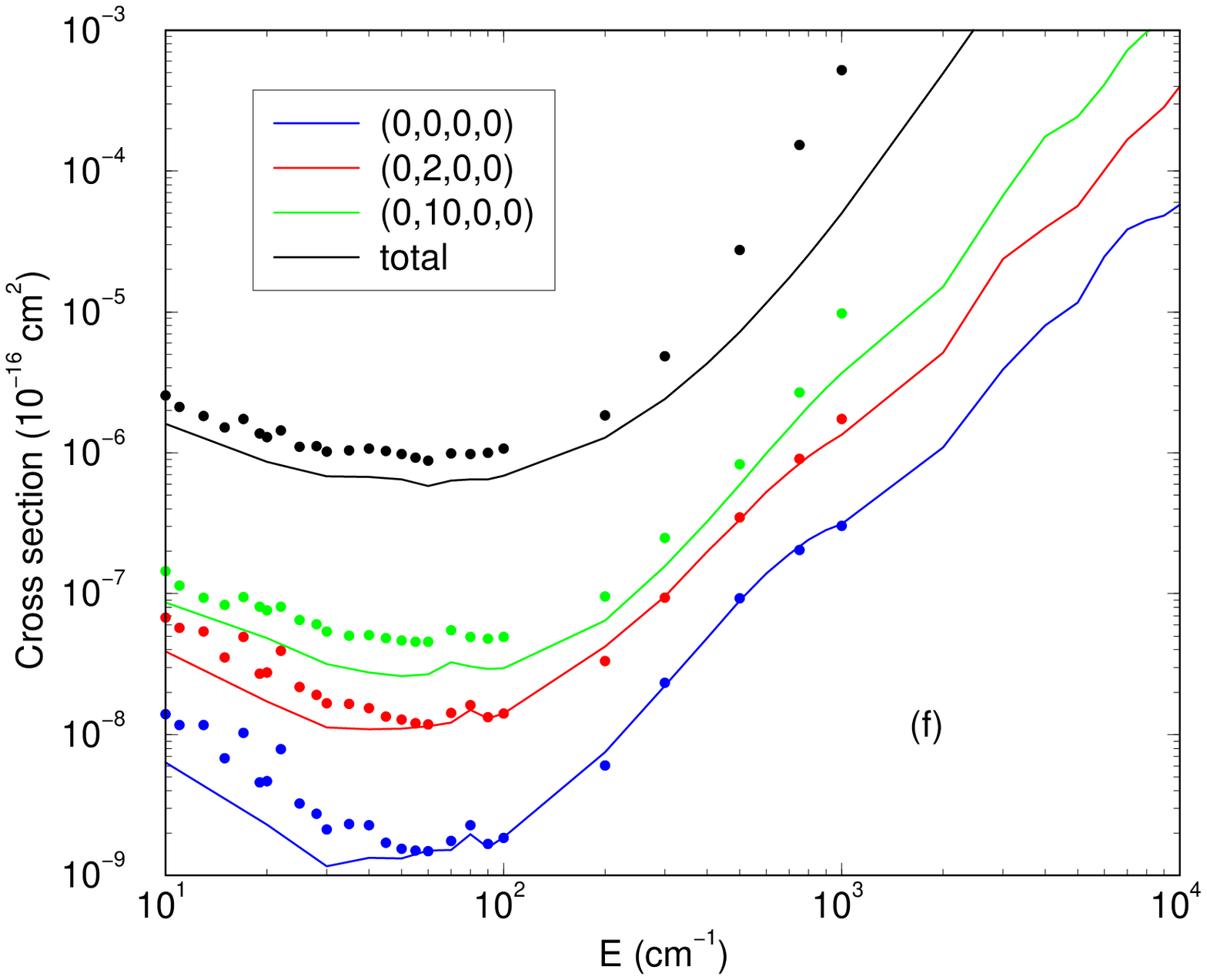}}
\caption{Vibrationally inelastic cross sections for transitions
between the initial state
(a) $(1,0,0,0)$,
(b) $(1,1,0,0)$,
(c) $(1,2,0,0)$,
(d) $(1,3,0,0)$
(e) $(1,4,0,0)$,
(f) $(1,5,0,0)$
and the final state indicated.
Solid lines are 5D-CS results and points are 6D-CC results \cite{yang}.}
\end{figure}

\newpage

\begin{figure}
\centerline{\epsfxsize=3in\epsfbox{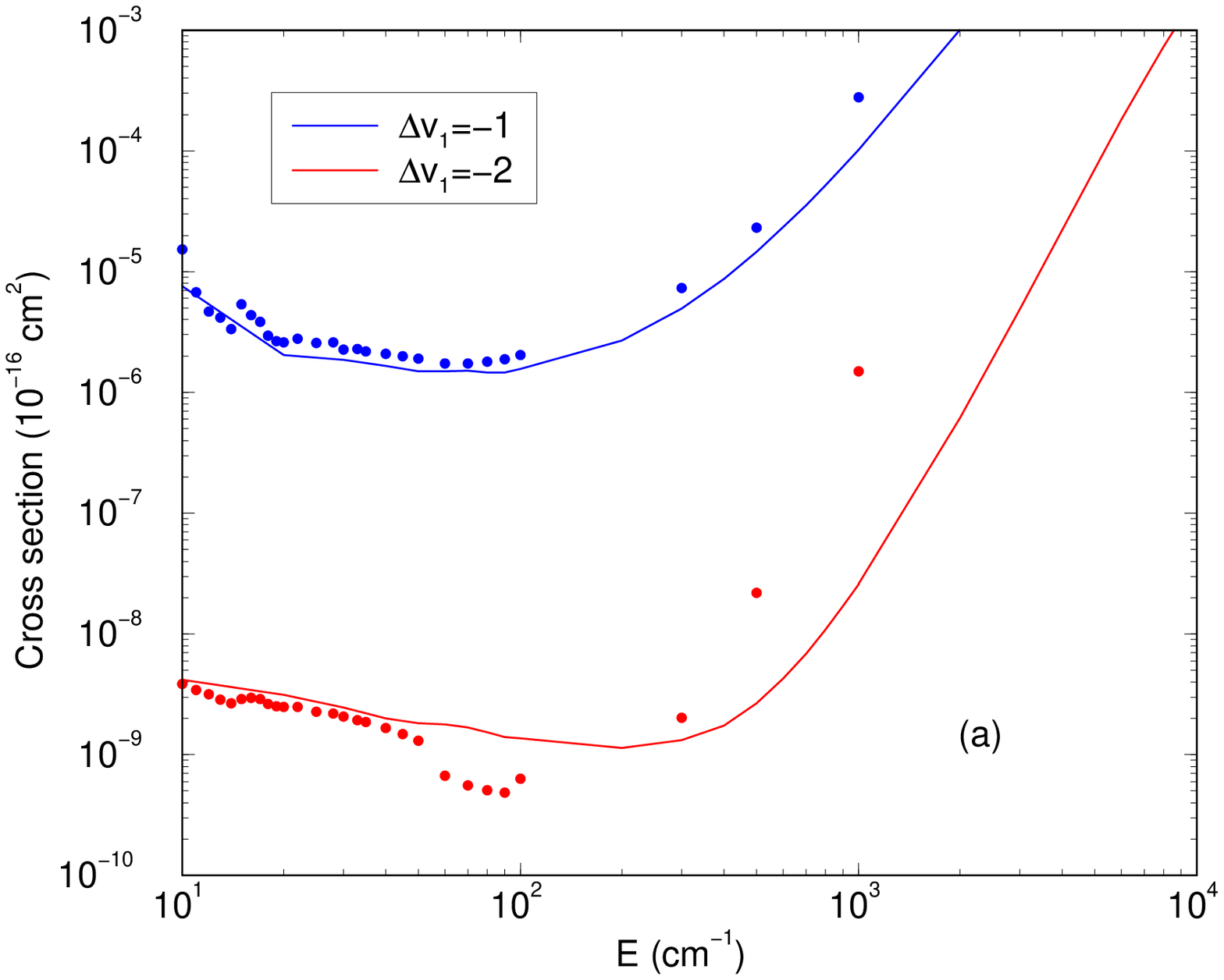}\hspace{.1in}\epsfxsize=3in\epsfbox{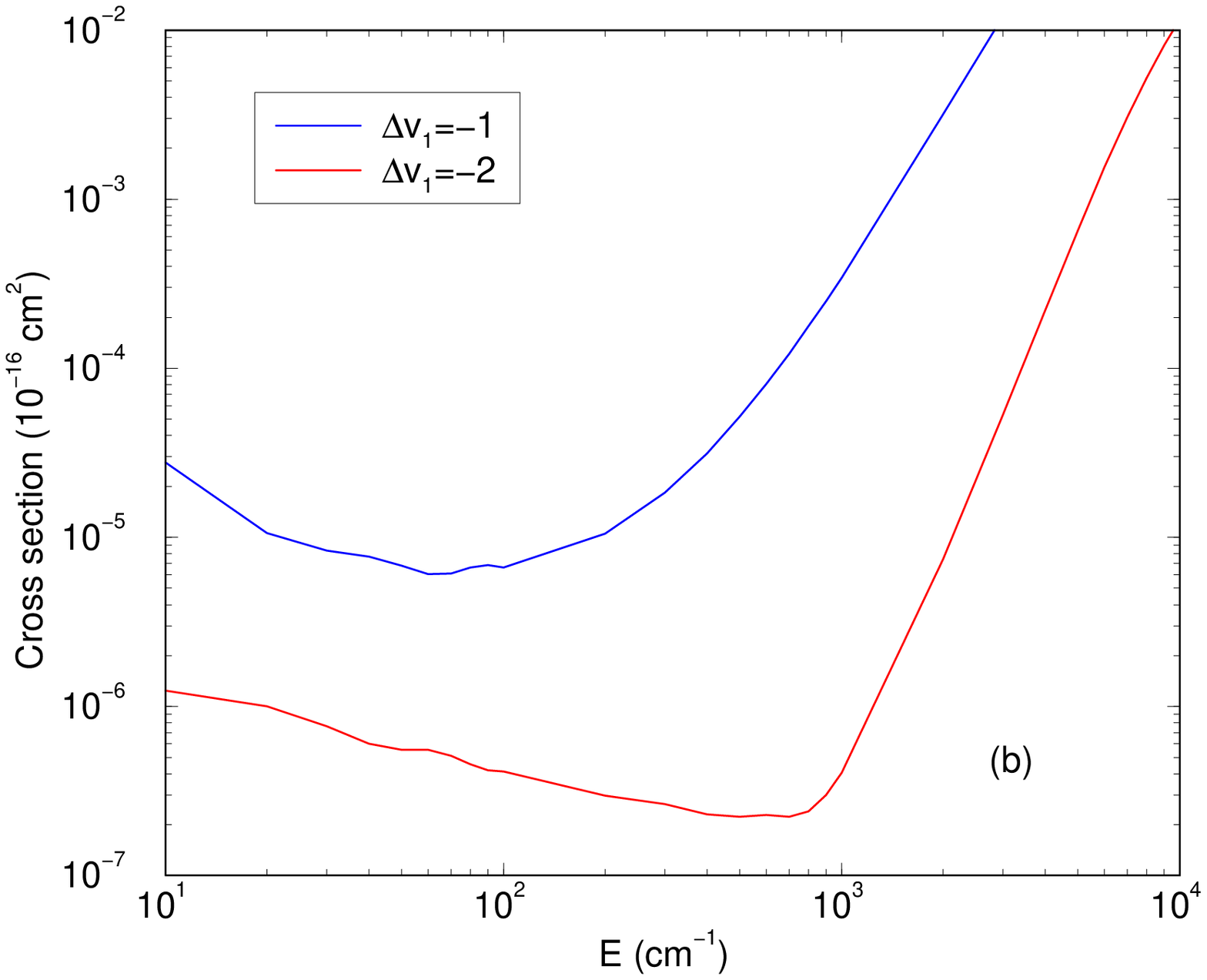}}
\caption{Vibrational quenching cross sections for (a) $(2,0,0,0)$ and (b) $(5,0,0,0)$ initial 
states.  Solid lines are 5D-CS results and points are 6D-CC results \cite{yang}.
The $\Delta v_1=-2$ transitions are at least one order of magnitude smaller
than $\Delta v_1=-1$ transitions for both sets of calculations.  }
\end{figure}

\newpage

\begin{figure}
\centerline{\epsfxsize=3in\epsfbox{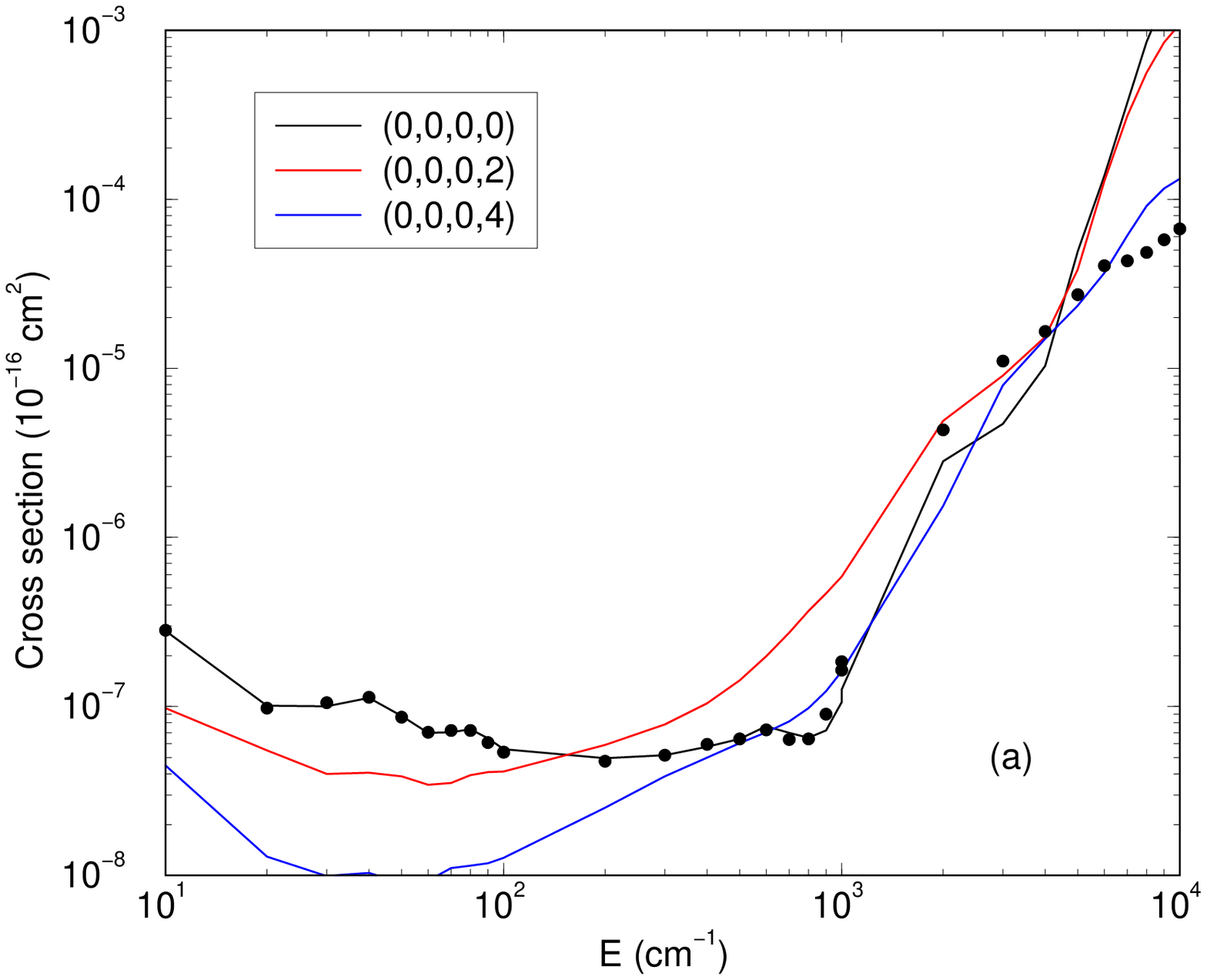}\hspace{.1in}\epsfxsize=3in\epsfbox{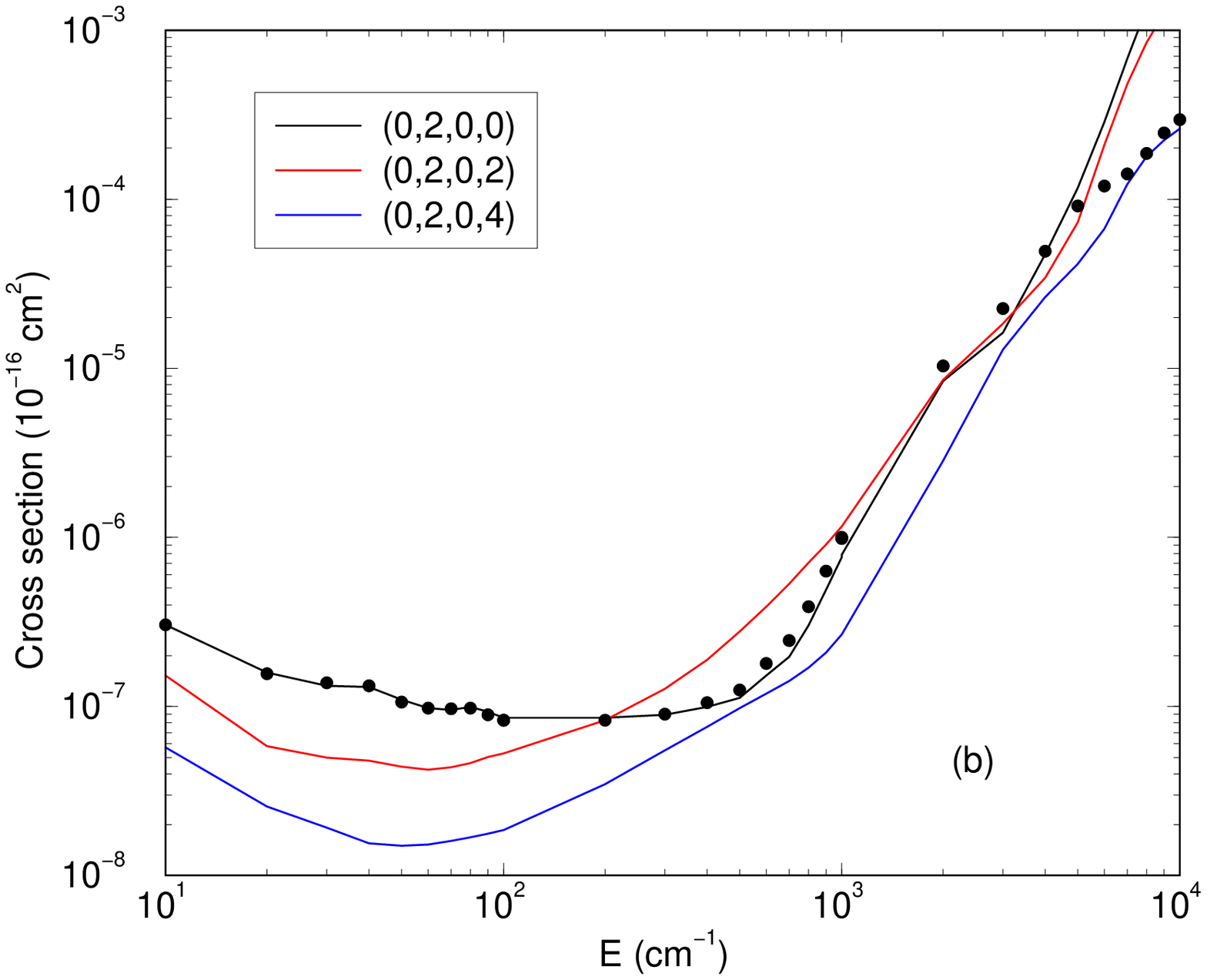}}
\centerline{\epsfxsize=3in\epsfbox{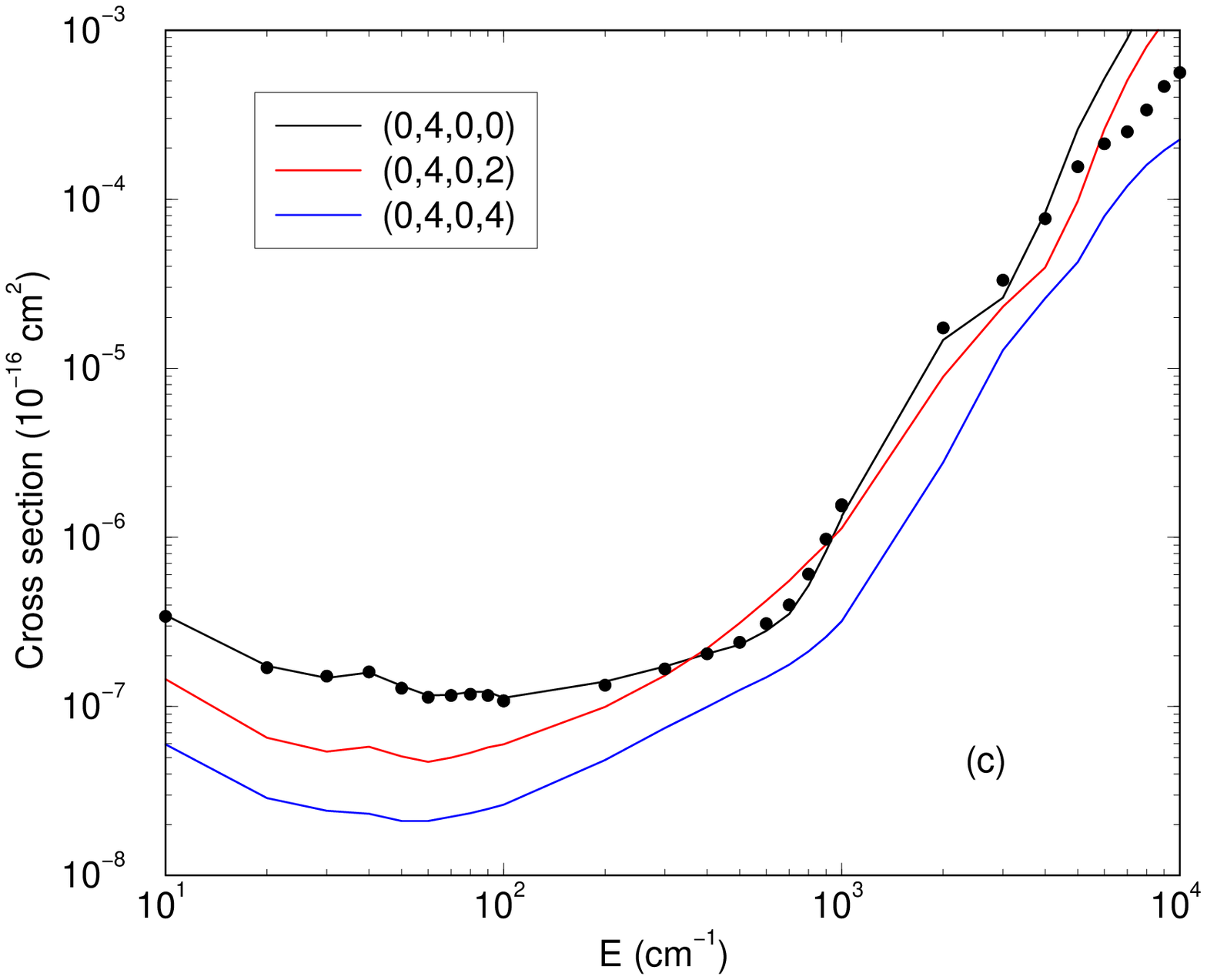}\hspace{.1in}\epsfxsize=3in\epsfbox{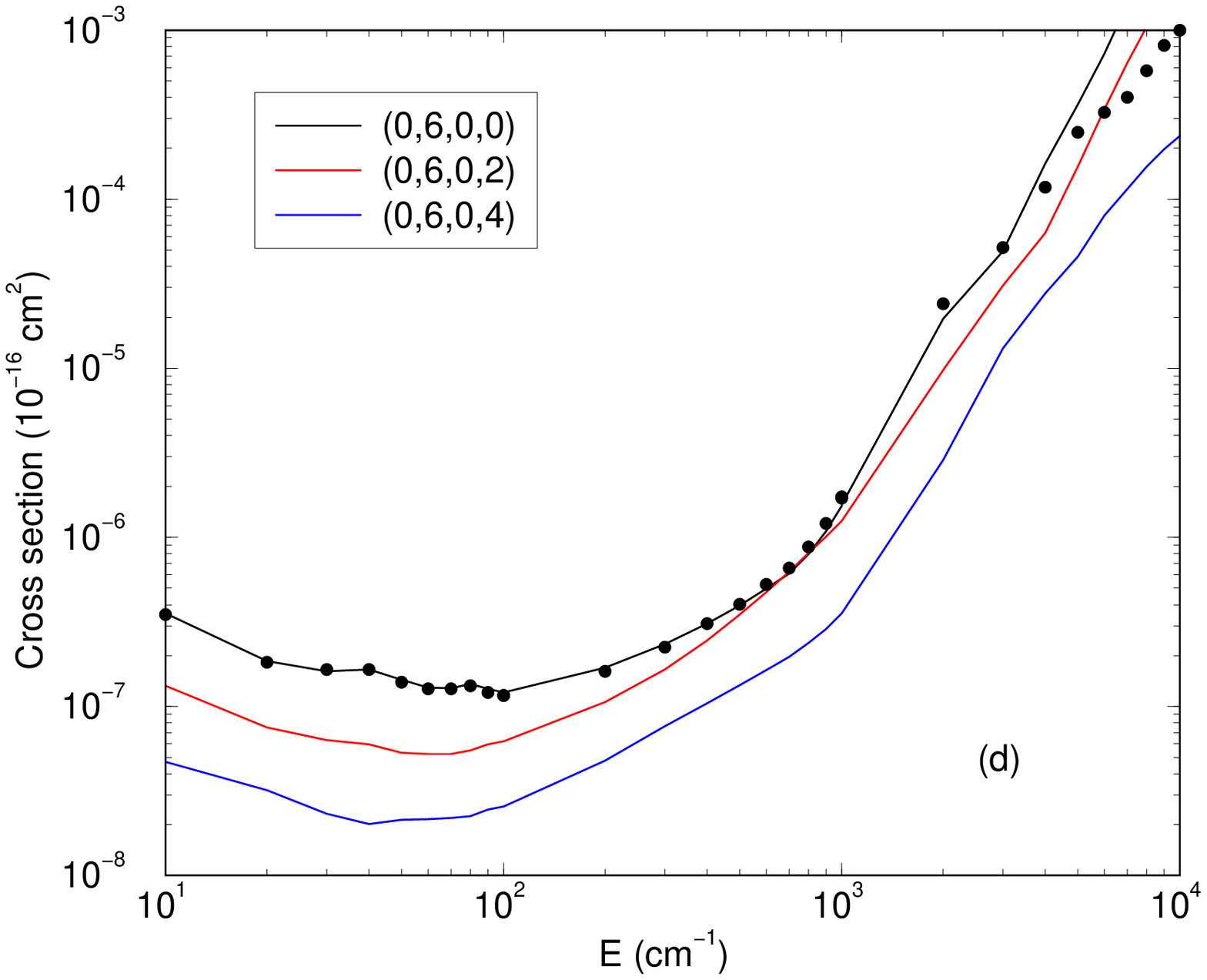}}
\centerline{\epsfxsize=3in\epsfbox{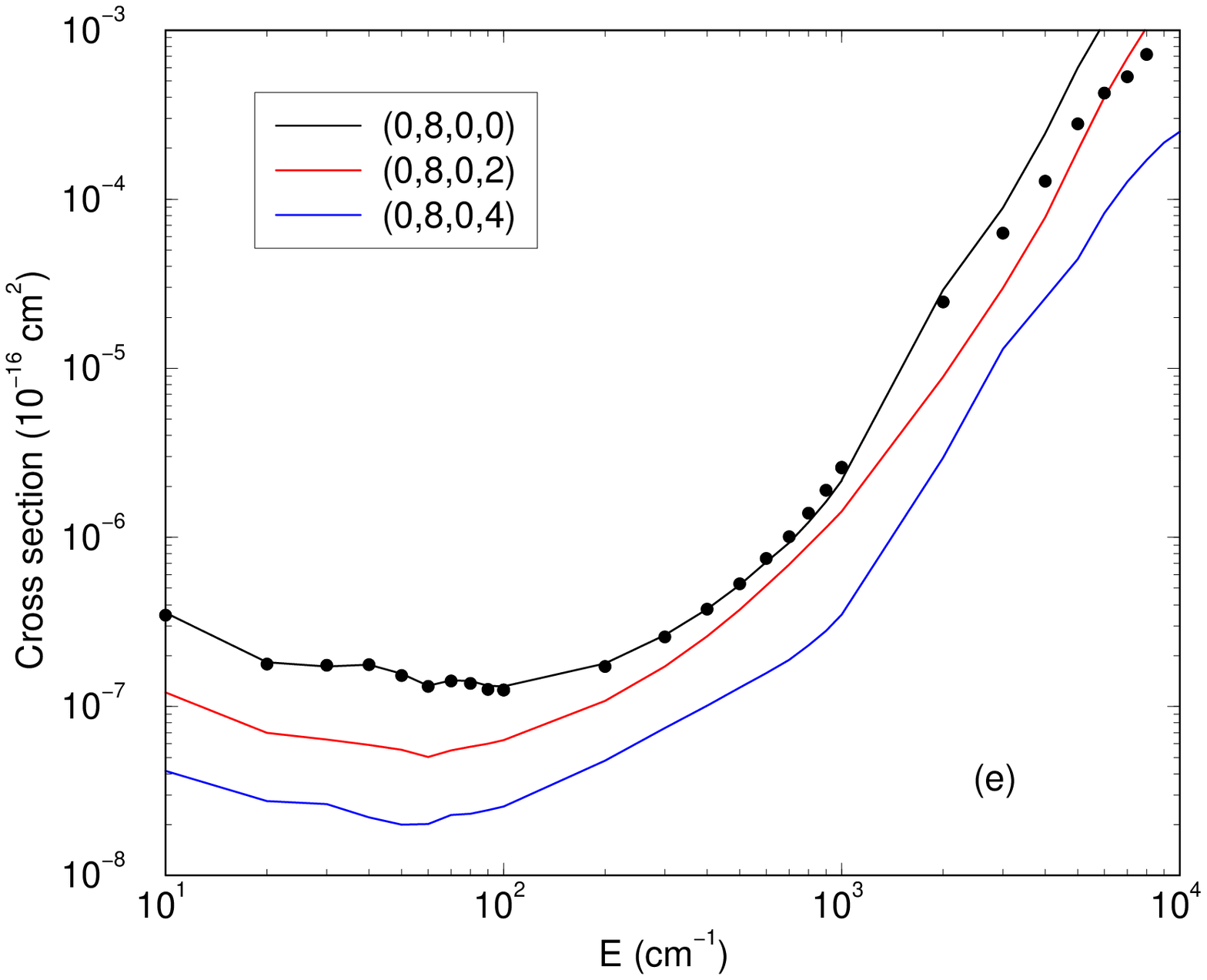}\hspace{.1in}\epsfxsize=3in\epsfbox{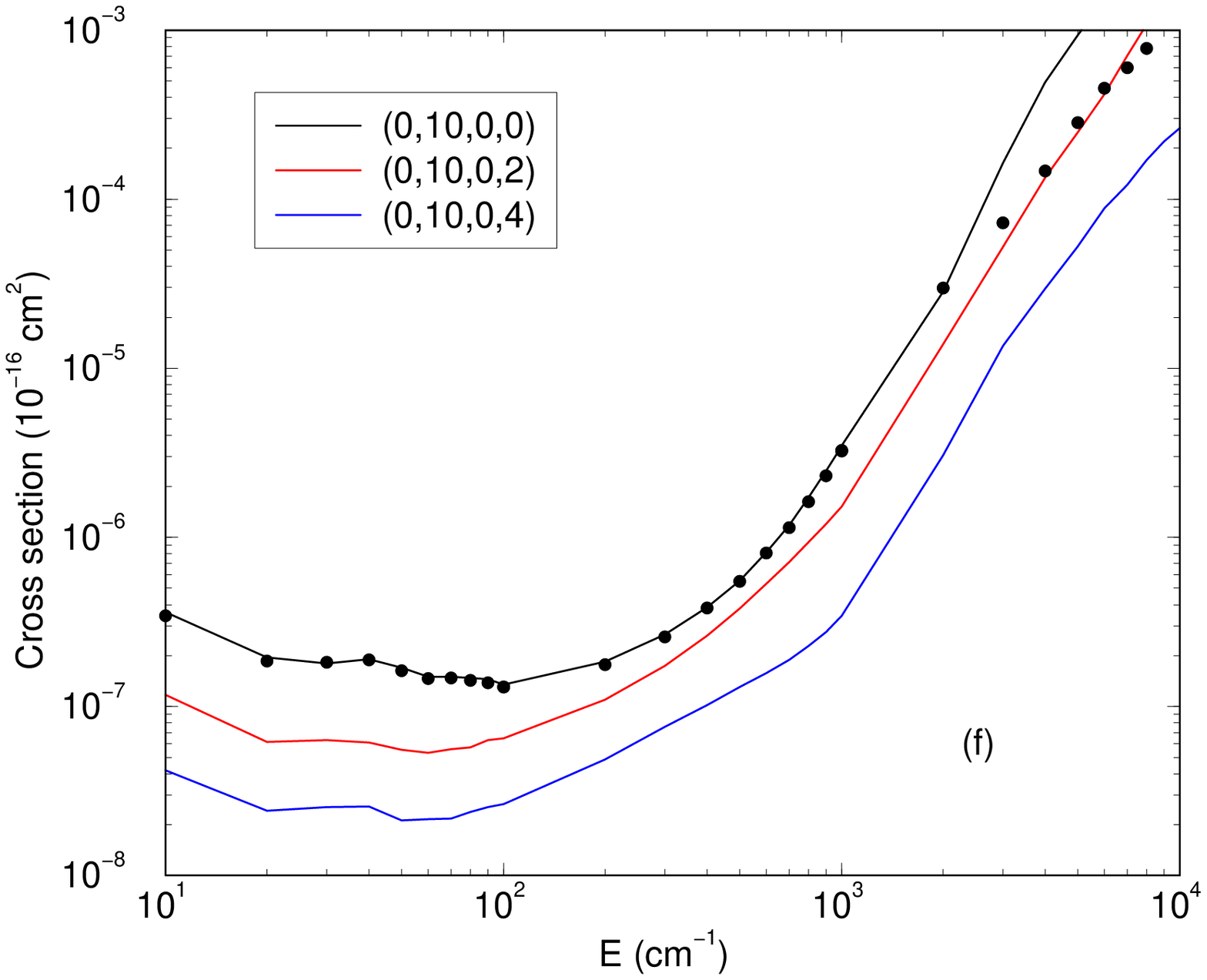}}
\caption{Vibrationally inelastic cross sections for transitions
between the initial state
(a) $(1,0,0,0)$,
(b) $(1,2,0,0)$,
(c) $(1,4,0,0)$,
(d) $(1,6,0,0)$
(e) $(1,8,0,0)$,
(f) $(1,10,0,0)$
and the final state indicated.
Solid lines are 5D-CS results which allow H$_2$ rotational transitions 
and points are 5D-CS results which restrict H$_2$ to be in the ground
rovibrational state.}
\end{figure}

\newpage

\begin{figure}
\centerline{\epsfxsize=3in\epsfbox{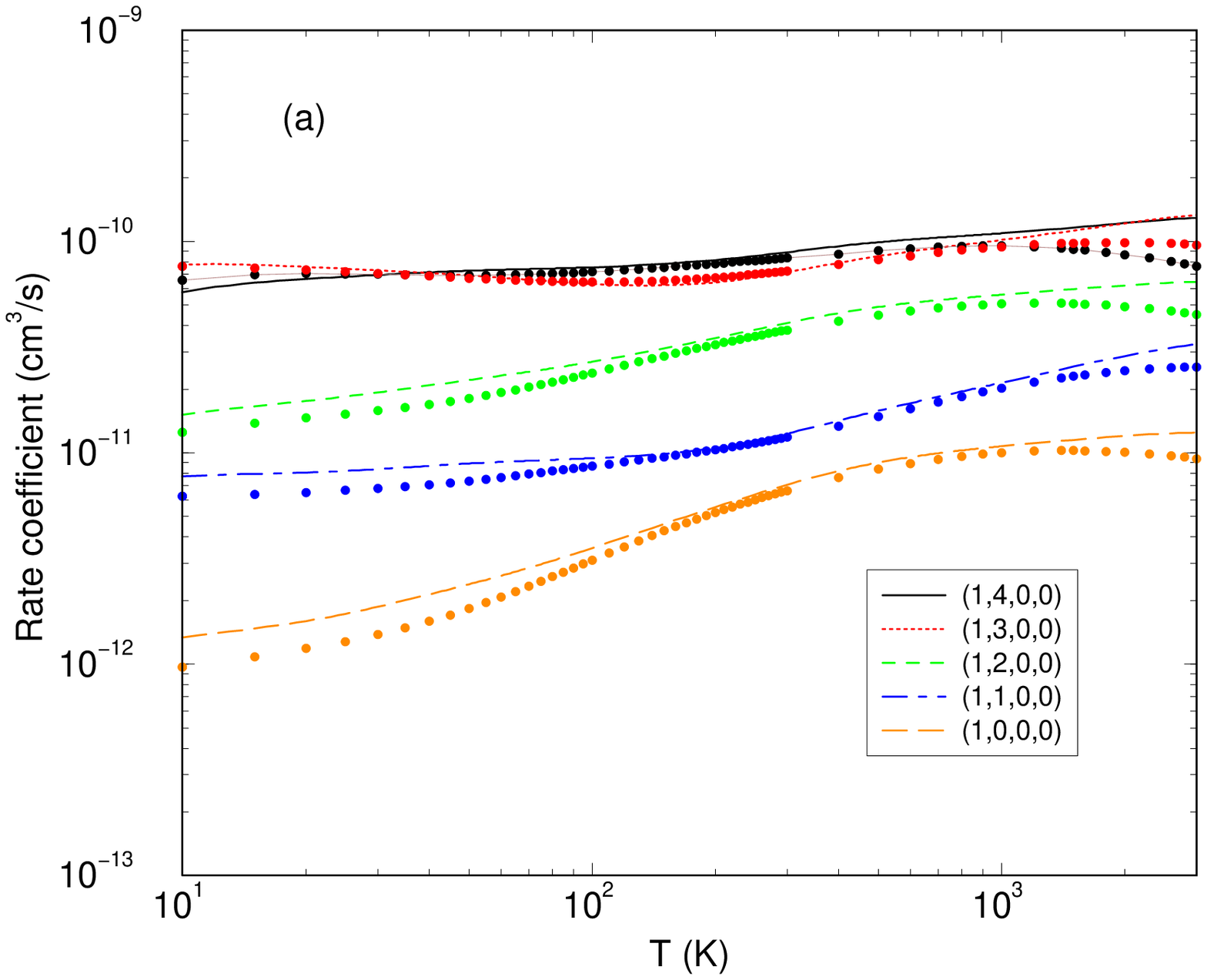}\hspace{.1in}\epsfxsize=3in\epsfbox{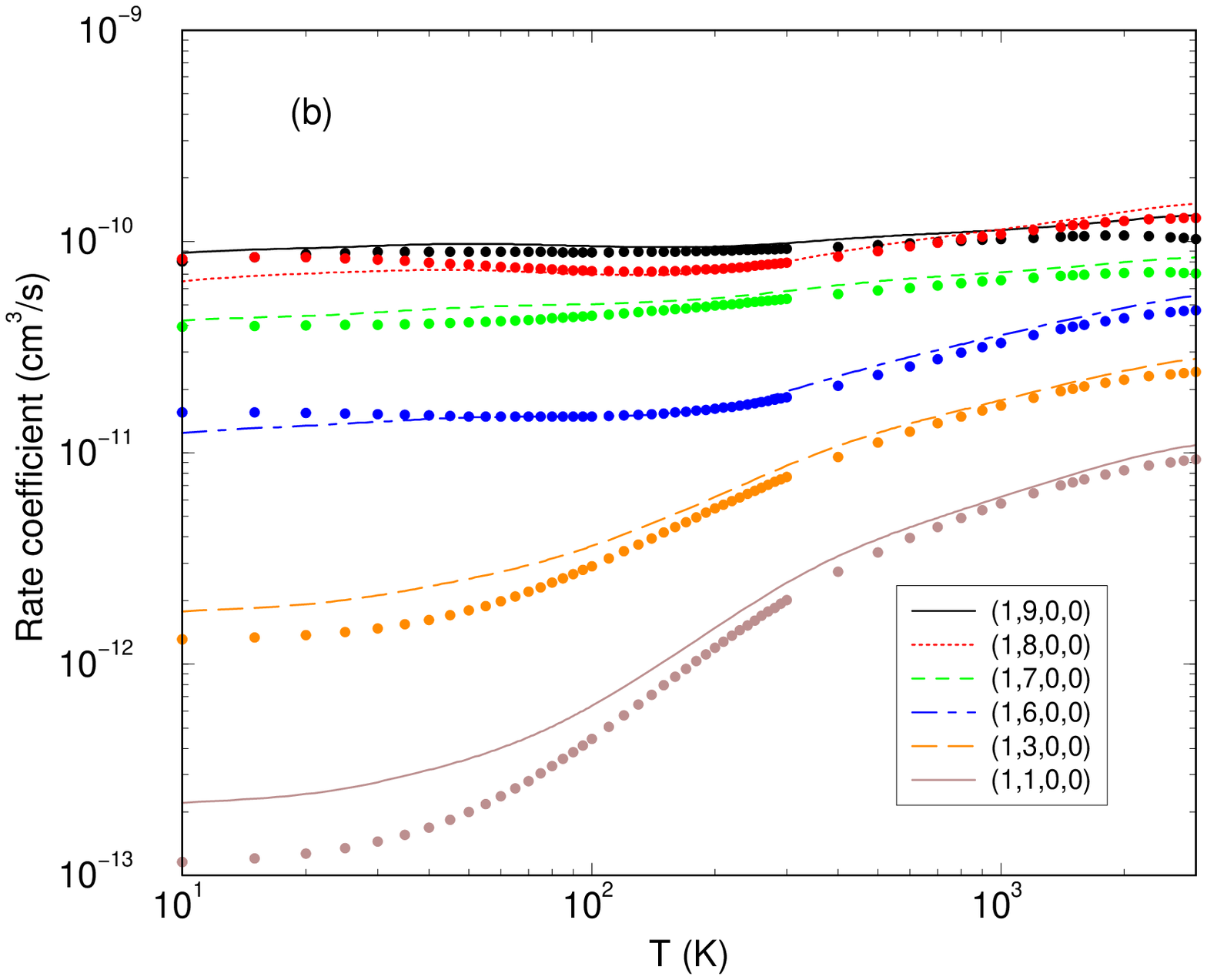}}
\centerline{\epsfxsize=3in\epsfbox{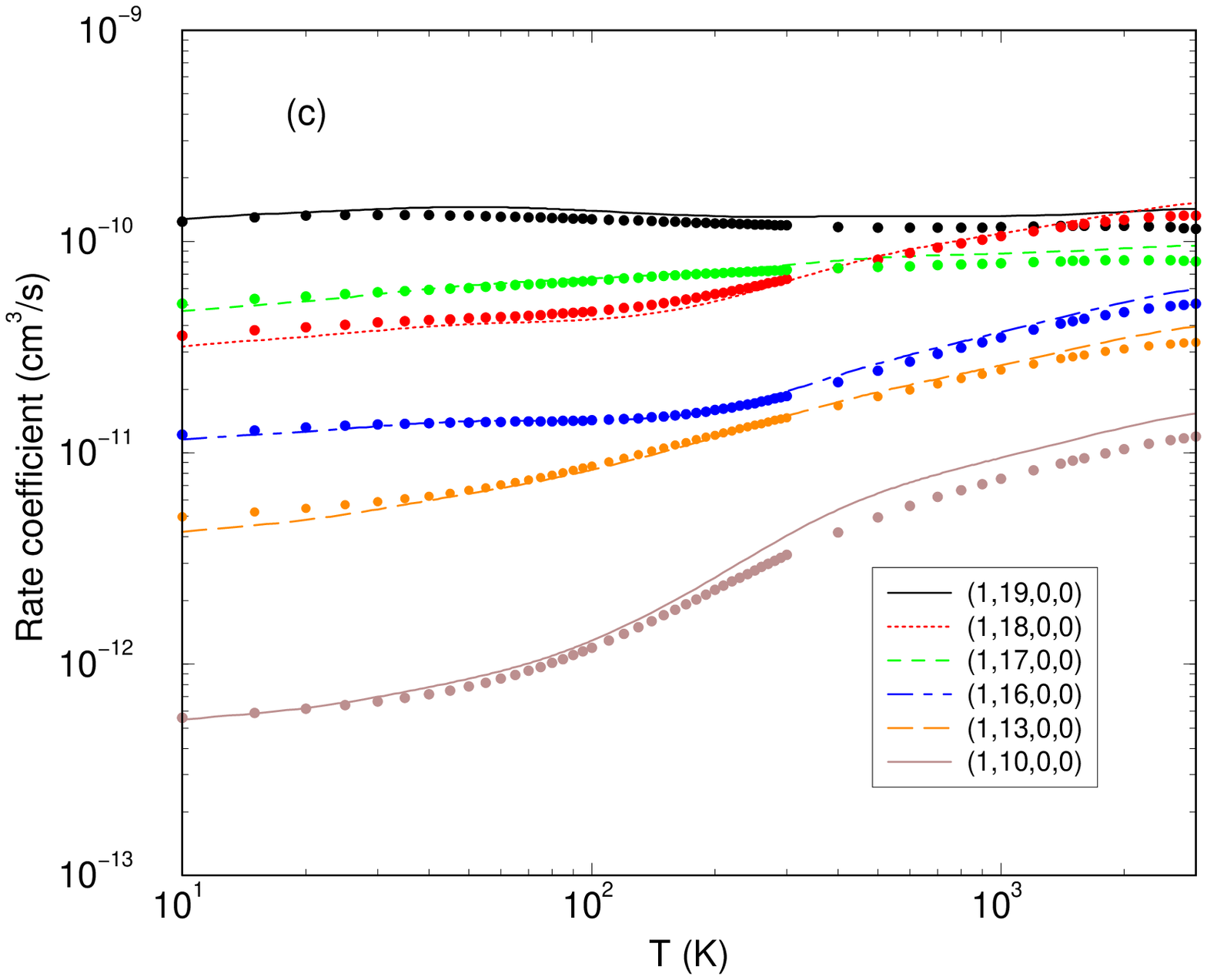}\hspace{.1in}\epsfxsize=3in\epsfbox{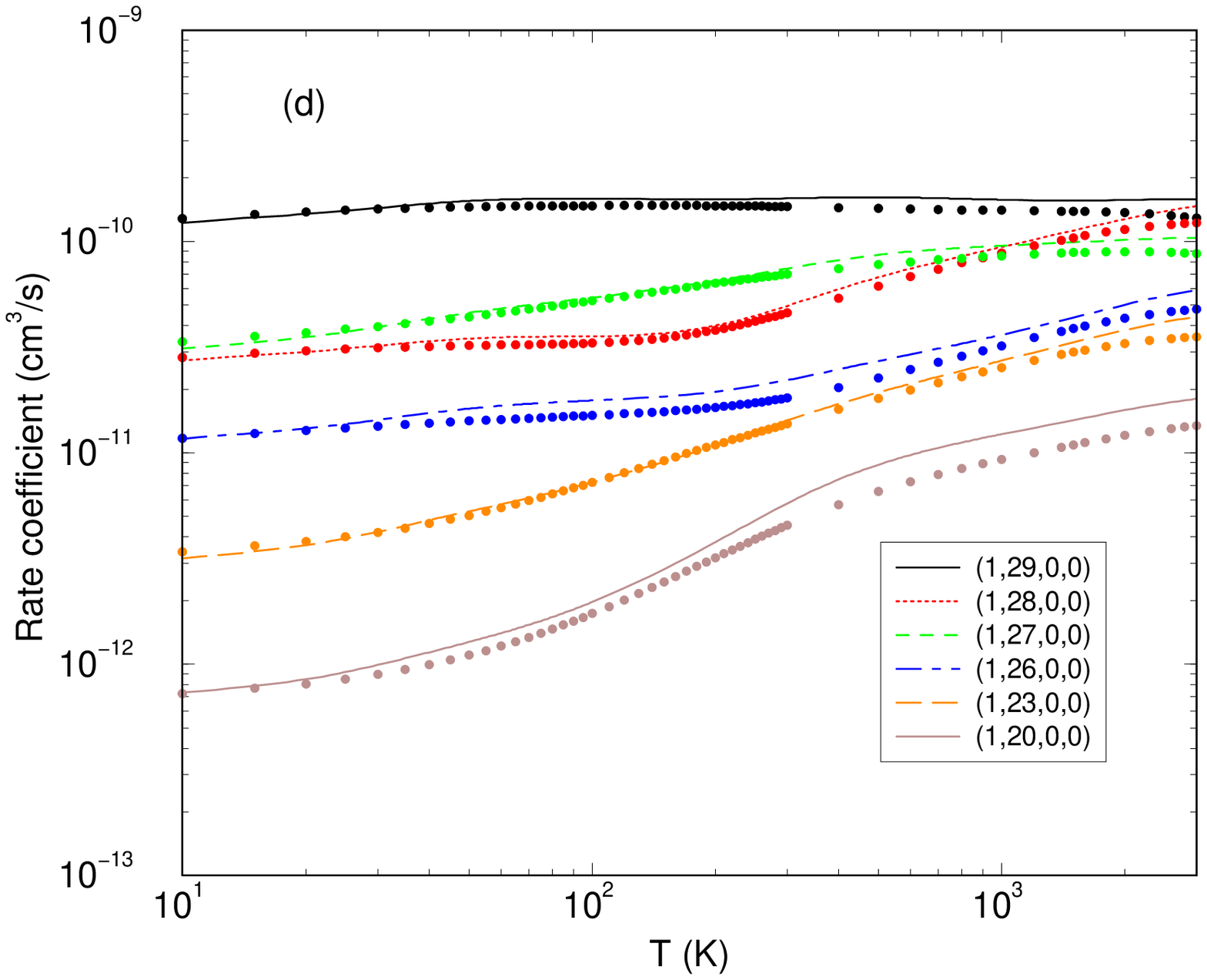}}
\caption{Rate coefficients for rotational transitions from initial state
(a) $(1,5,0,0)$, (b) $(1,10,0,0)$, (c) $(1,20,0,0)$ and (d) $(1,30,0,0)$. 
Also shown (colored points) are rigid rotor results 
reported previously \cite{rotor} for $(0,j_1,0,0)$.
}
\end{figure}

\newpage

\begin{figure}
\centerline{\epsfxsize=3in\epsfbox{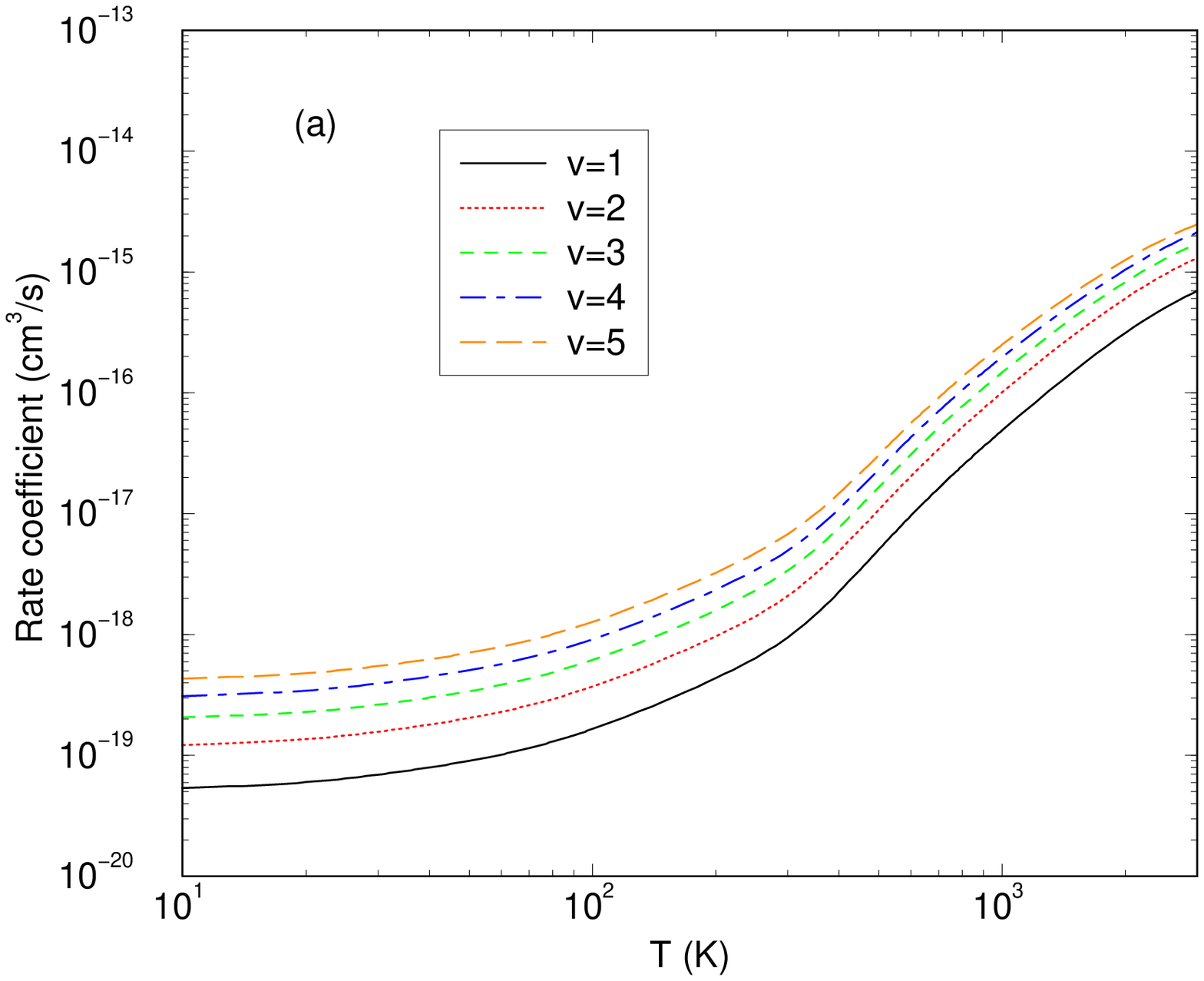}\hspace{.1in}\epsfxsize=3in\epsfbox{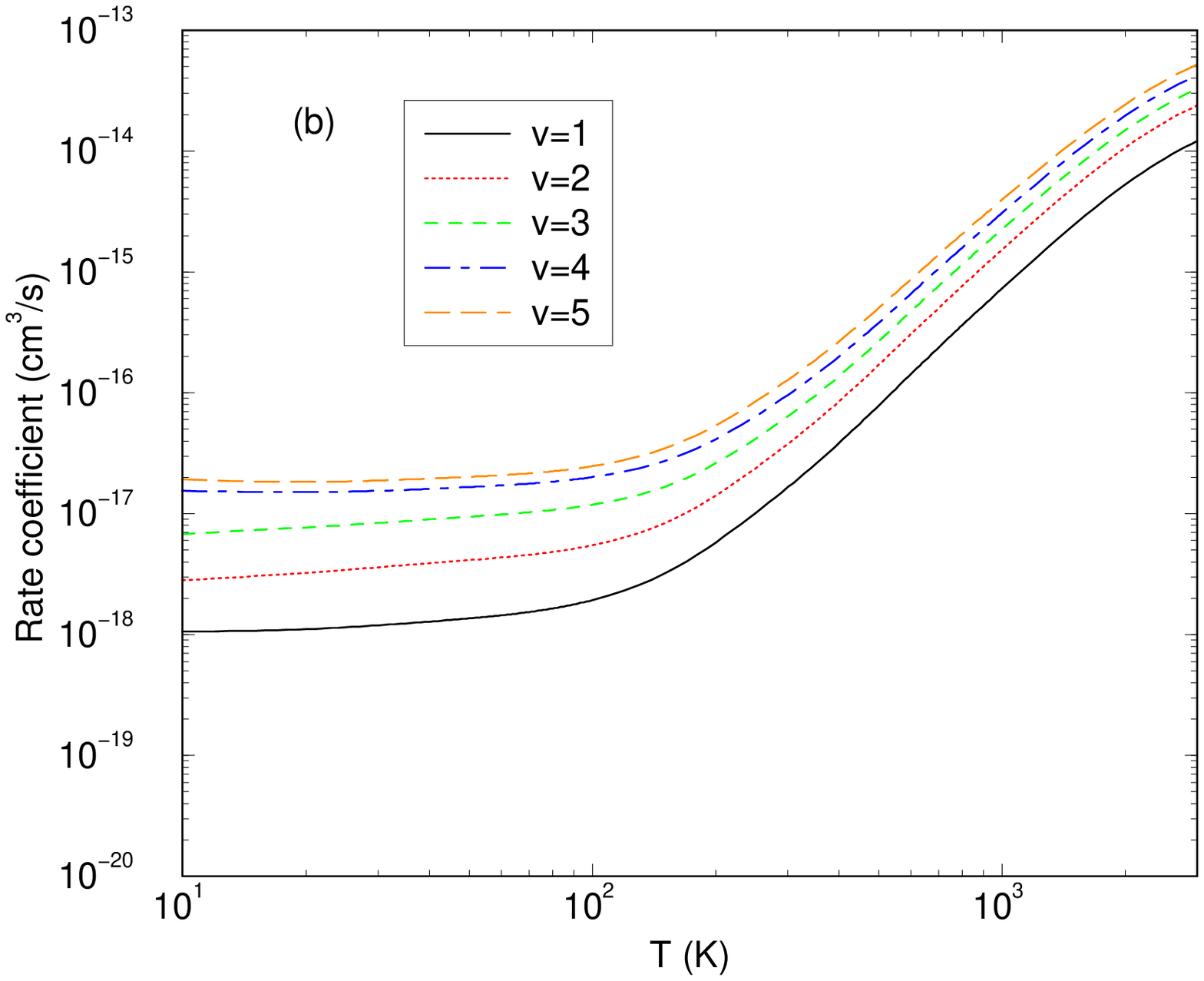}}
\centerline{\epsfxsize=3in\epsfbox{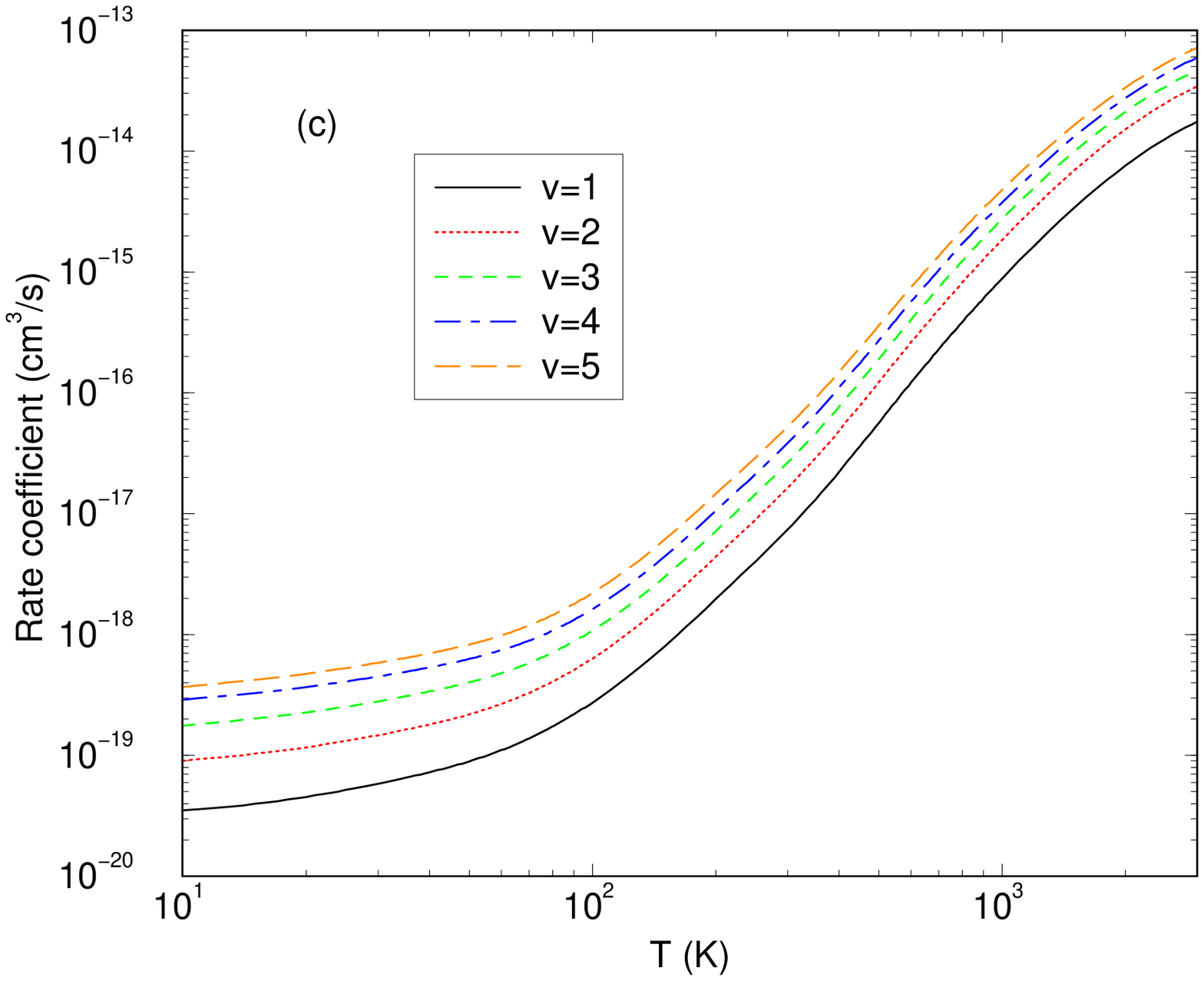}\hspace{.1in}\epsfxsize=3in\epsfbox{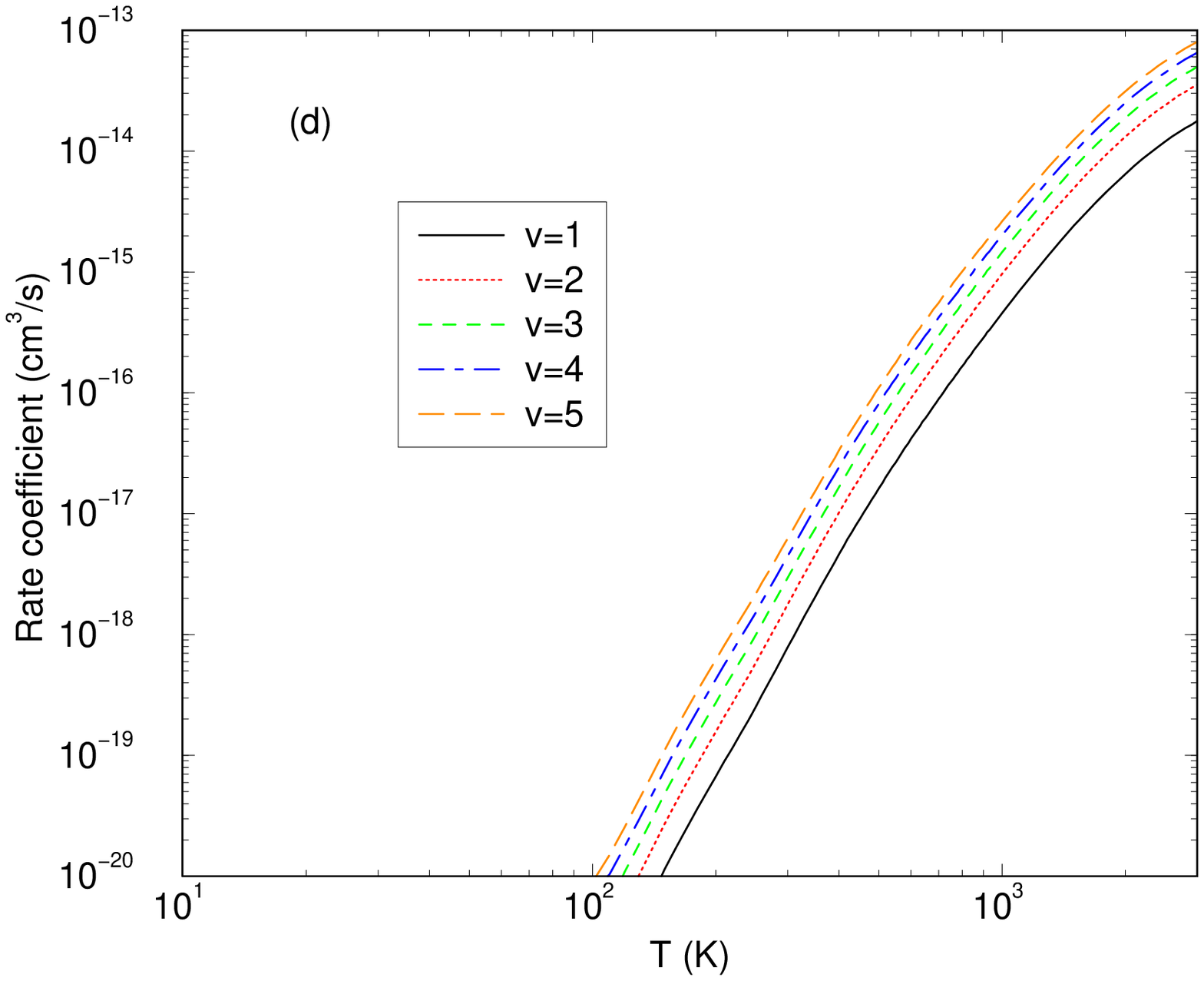}}
\caption{Rate coefficients for transitions from $(v,10,0,0)$ to (a) $(v-1,0,0,0)$,
(b) $(v-1,10,0,0)$, (c) $(v-1,20,0,0)$, and (d) $(v-1,30,0,0)$ obtained from
the ``frozen-H$_2$" basis set.
}
\end{figure}

\newpage

\begin{figure}
\centerline{\epsfxsize=3in\epsfbox{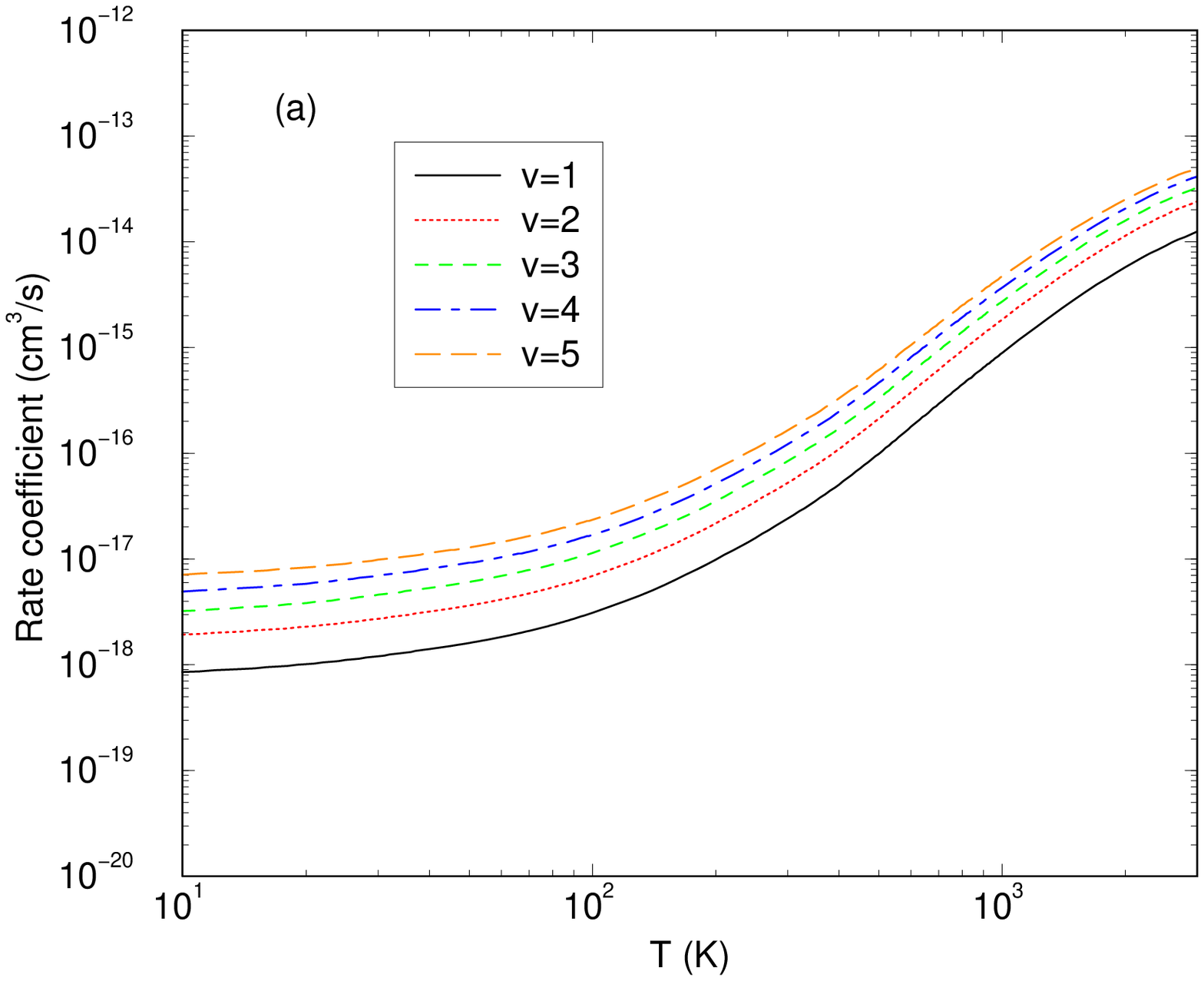}\hspace{.1in}\epsfxsize=3in\epsfbox{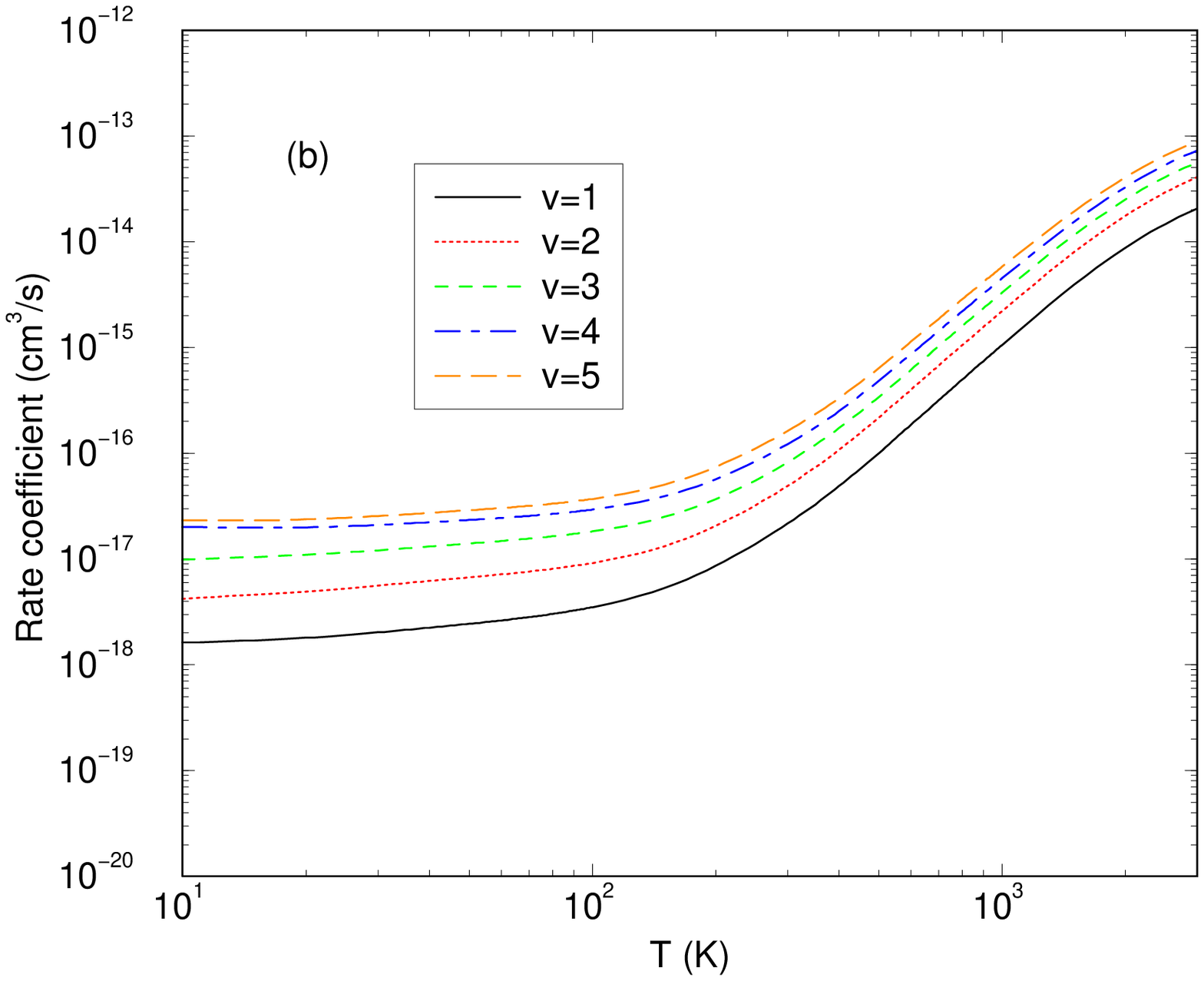}}
\centerline{\epsfxsize=3in\epsfbox{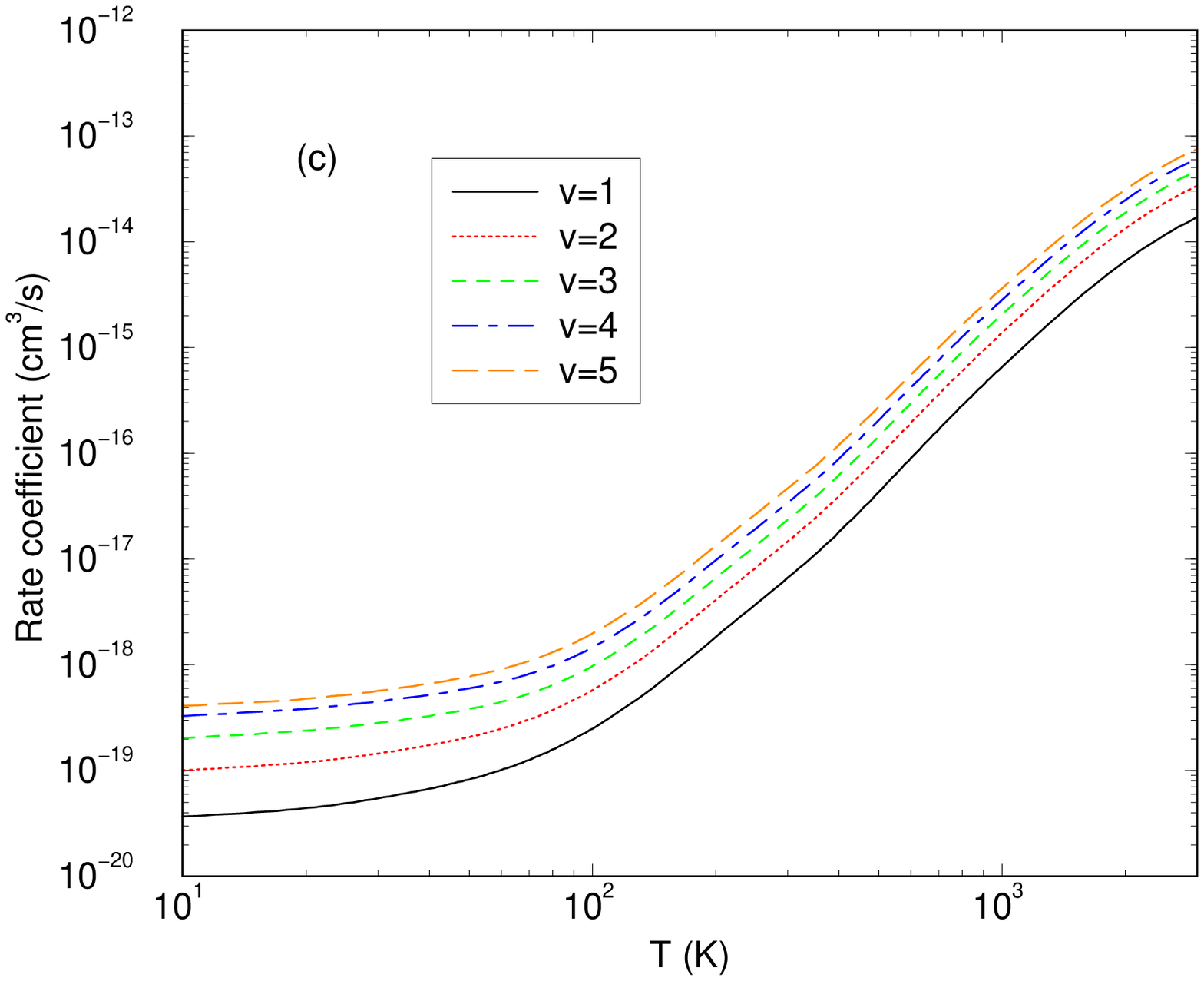}\hspace{.1in}\epsfxsize=3in\epsfbox{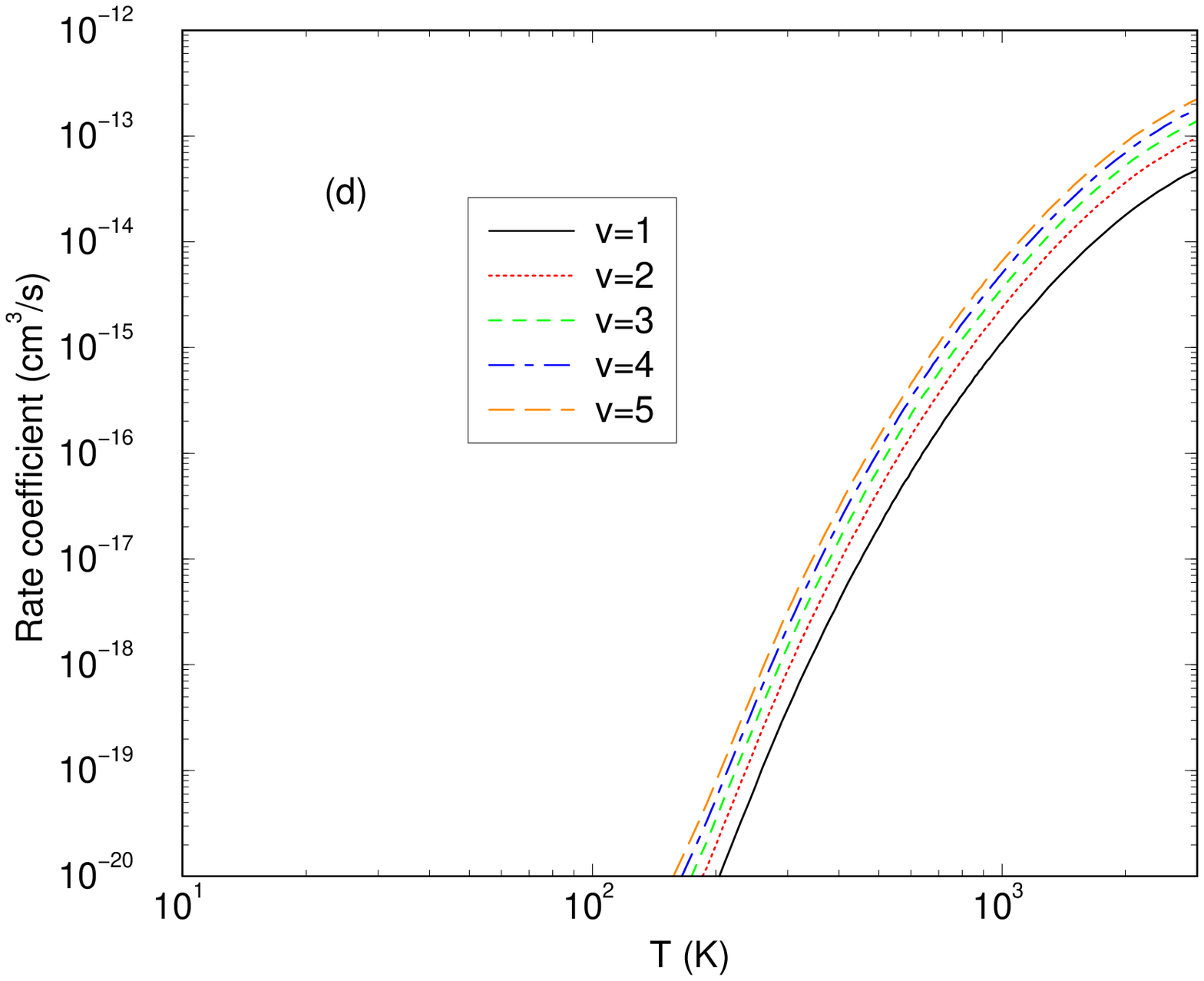}}
\caption{Rate coefficients for transitions from $(v,20,0,0)$ to (a) $(v-1,10,0,0)$,
(b) $(v-1,20,0,0)$, (c) $(v-1,30,0,0)$, and (d) $(v-1,40,0,0)$ obtained from
the ``frozen-H$_2$" basis set.
}
\end{figure}

\newpage

\begin{figure}
\centerline{\epsfxsize=3in\epsfbox{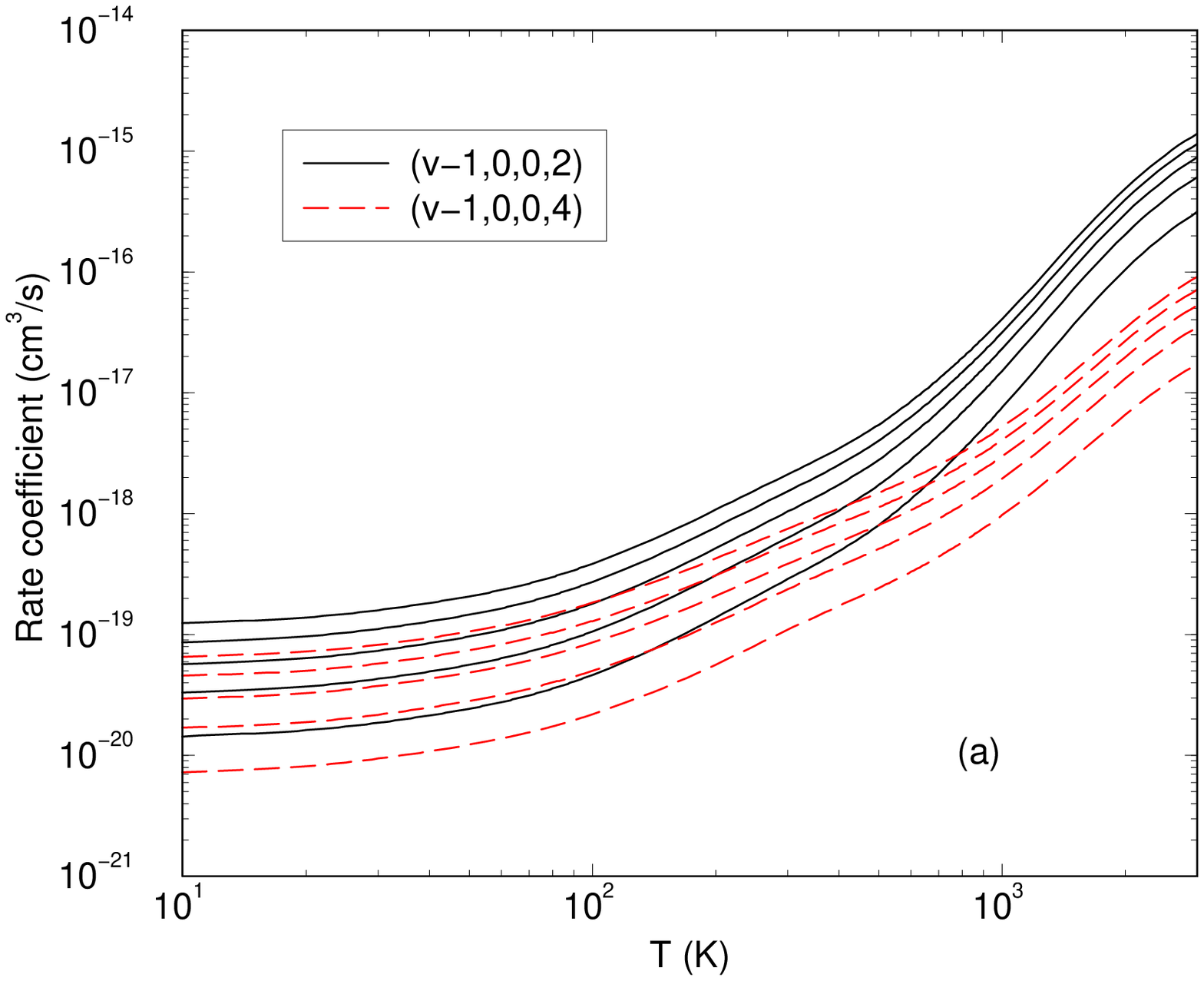}\hspace{.1in}\epsfxsize=3in\epsfbox{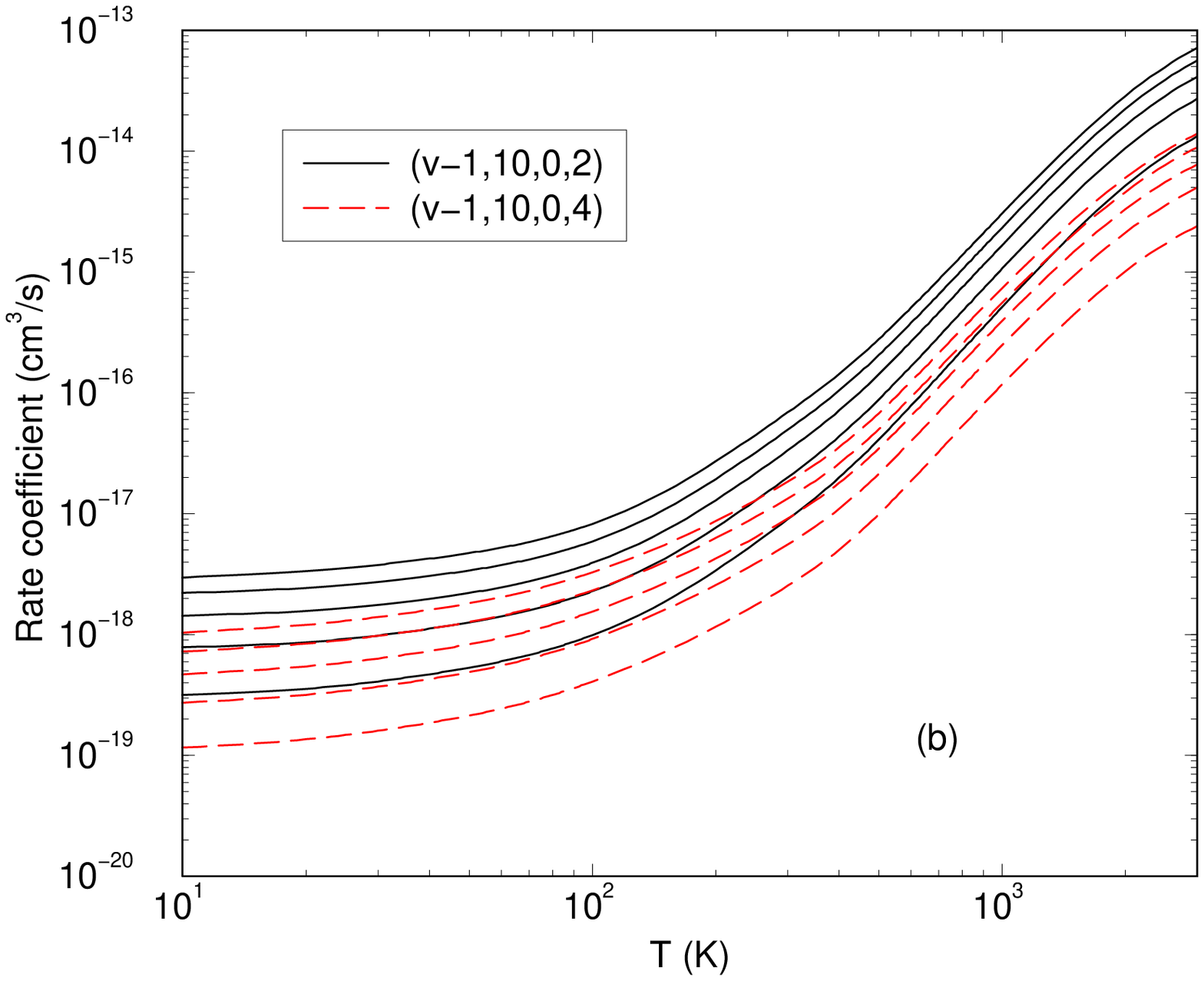}}
\centerline{\epsfxsize=3in\epsfbox{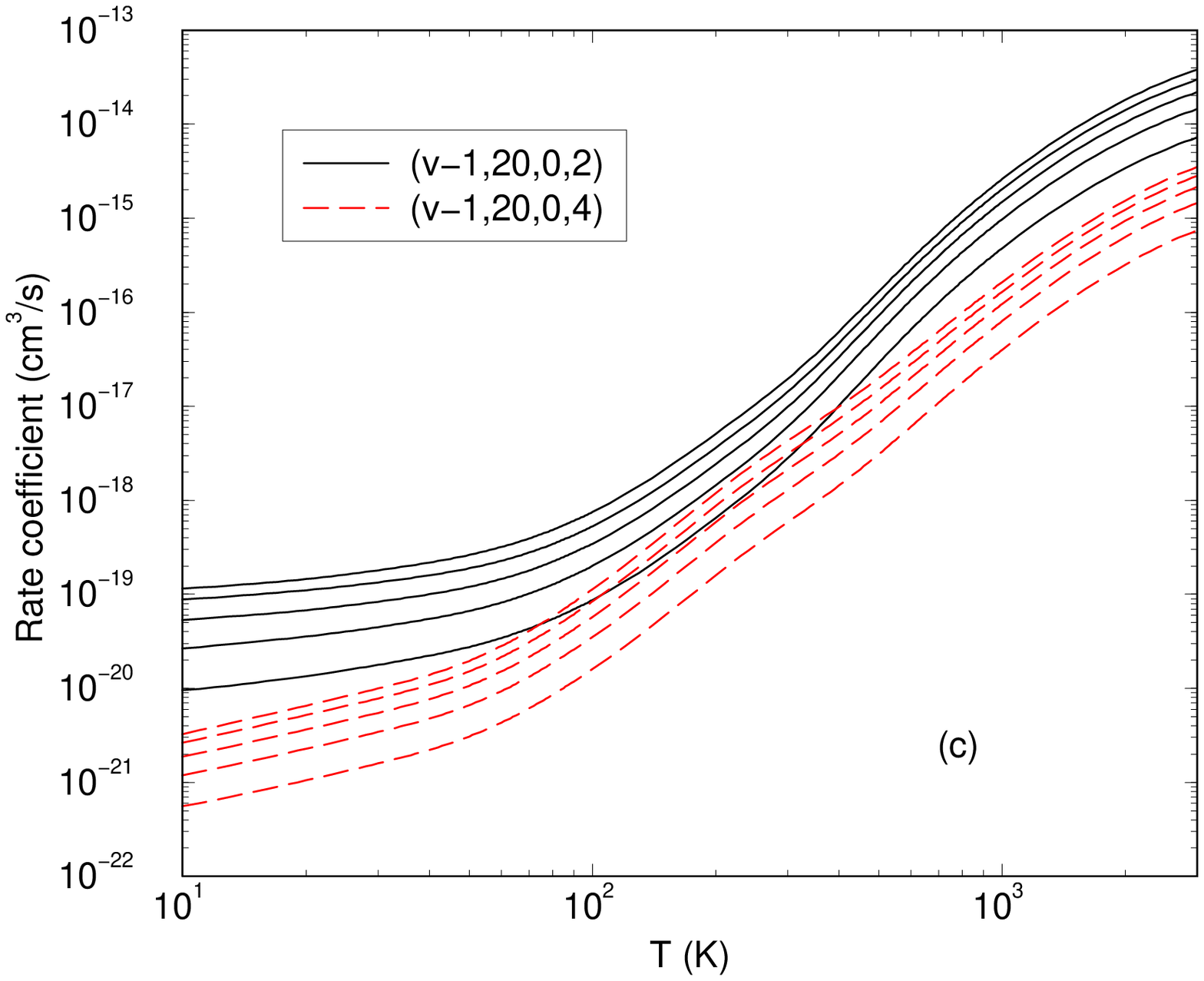}\hspace{.1in}\epsfxsize=3in\epsfbox{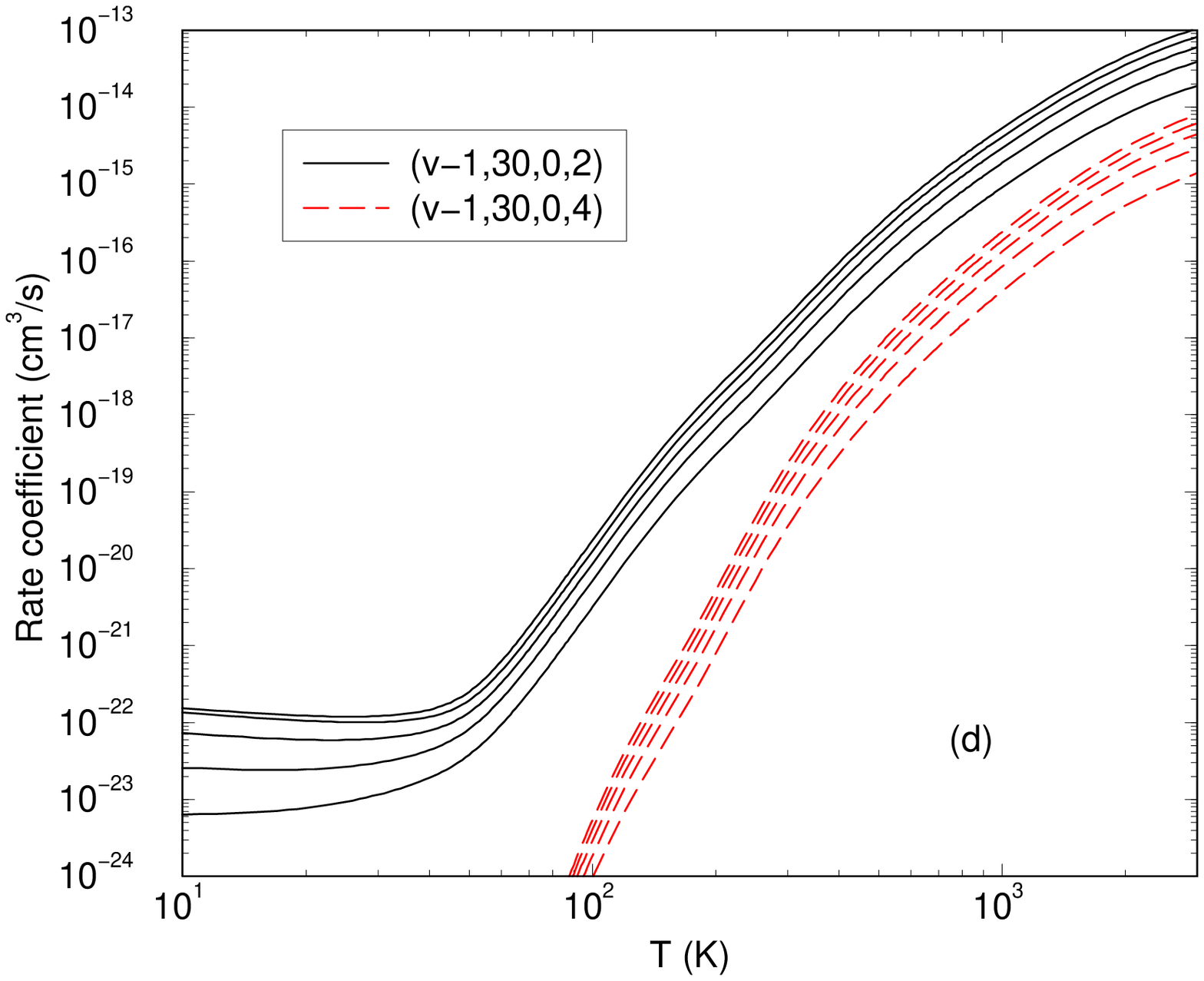}}
\caption{Rate coefficients for transitions from $(v,10,0,0)$ to (a) $(v-1,0,0,j)$,
(b) $(v-1,10,0,j)$, (c) $(v-1,20,0,j)$, and (d) $(v-1,30,0,j)$ obtained from
the ``flexible-H$_2$" basis set. The solid black curves correspond to 
H$_2(0,2)$ final states, and the dashed red curves to H$_2(0,4)$ final states.
Both sets of curves increase uniformly with $v$ for $v=1-5$.
}
\end{figure}

\newpage

\begin{figure}
\centerline{\epsfxsize=3in\epsfbox{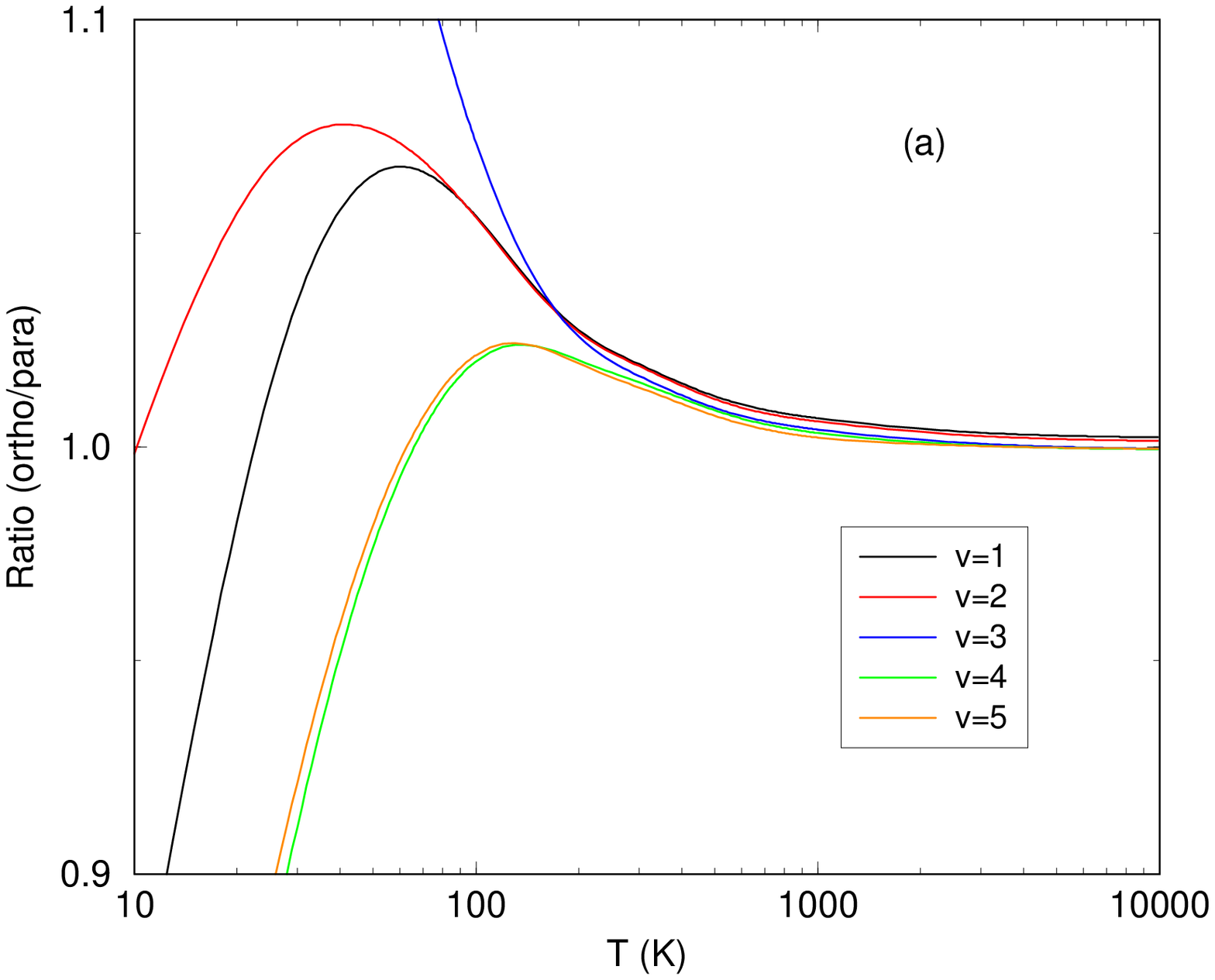}\hspace{.1in}\epsfxsize=3in\epsfbox{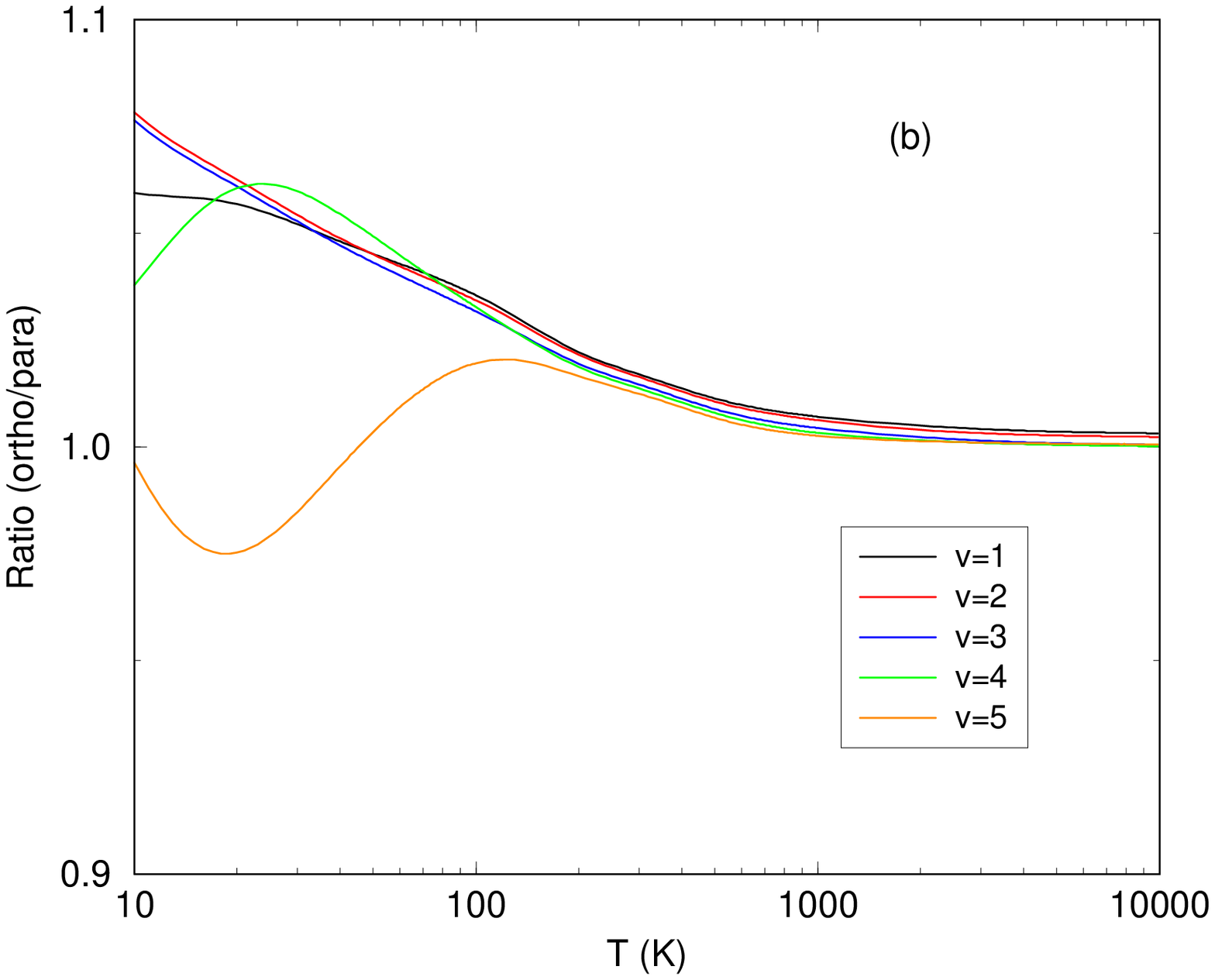}}
\centerline{\epsfxsize=3in\epsfbox{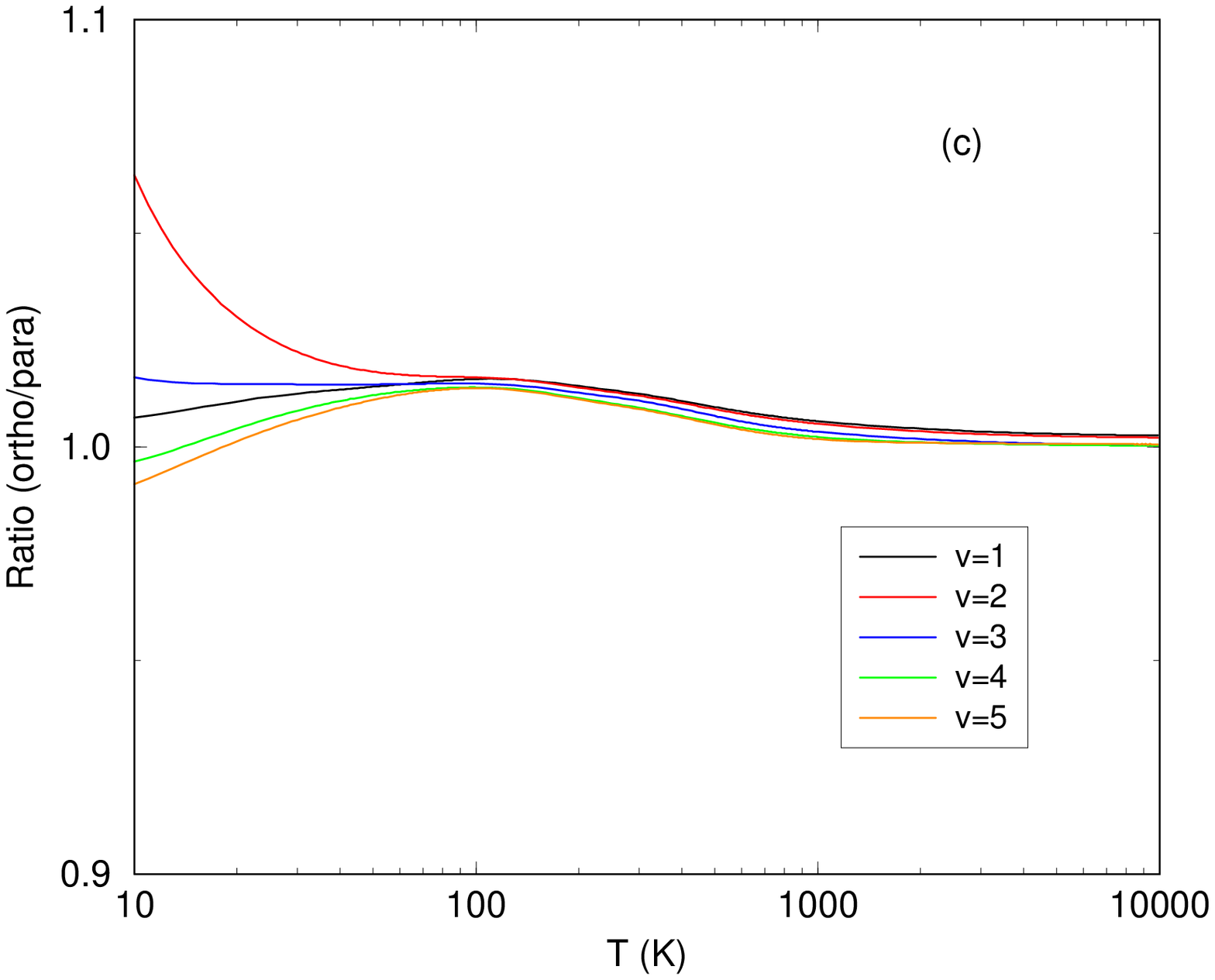}\hspace{.1in}\epsfxsize=3in\epsfbox{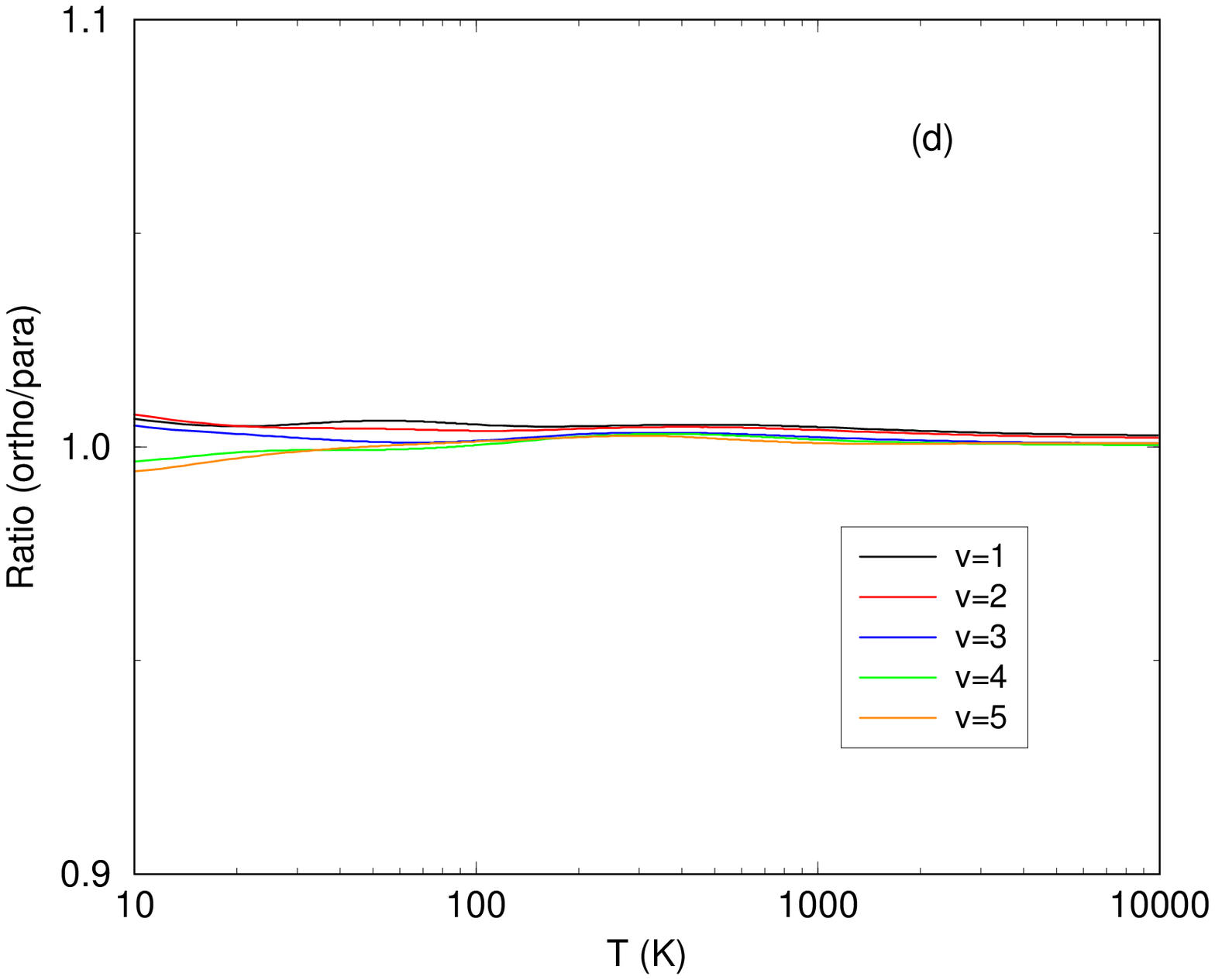}}
\caption{Ratio of vibrational quenching rate coefficients for collision with ortho-H$_2$ and para-H$_2$.
The panels correspond to CO in the initial state (a) $(v,0)$, (b) $(v,5)$, (c) $(v,10)$, and (d) $(v,20)$
making a transition to the final state $(v-1)$ summed over all final rotational levels. 
}
\end{figure}

\end{document}